\documentclass[12pt, titlepage]{article}\usepackage[]{graphicx}\usepackage[]{color}
\makeatletter
\def\maxwidth{ %
  \ifdim\Gin@nat@width>\linewidth
    \linewidth
  \else
    \Gin@nat@width
  \fi
}
\makeatother

\definecolor{fgcolor}{rgb}{0.345, 0.345, 0.345}

\usepackage{framed}
\makeatletter
 {\par\unskip\endMakeFramed%
 \at@end@of@kframe}
\makeatother

\definecolor{shadecolor}{rgb}{.97, .97, .97}
\definecolor{messagecolor}{rgb}{0, 0, 0}
\definecolor{warningcolor}{rgb}{1, 0, 1}
\definecolor{errorcolor}{rgb}{1, 0, 0}

\usepackage{alltt}

\usepackage{graphicx, ams, amsmath, amssymb, natbib, setspace}
\usepackage{float}
\usepackage{multirow}
\newcommand{\bx}{\ensuremath{\mathbf{x}}}
\newcommand{\bz}{\ensuremath{\mathbf{z}}}

\newcommand{\bs}{\ensuremath{\mathbf{s}}}
\newcommand{\bu}{\ensuremath{\mathbf{u}}}
\newcommand{\cS}{\ensuremath{\mathcal{S}}}

\def\bc{\mathbf{c}}

\def\bs{\mathbf{s}}
\def\bu{\mathbf{u}}

\def\by{\mathbf{y}}
\def\bz{\mathbf{z}}

\def\bW{\mathbf{W}}
\def\bX{\mathbf{X}}
\def\bY{\mathbf{Y}}
\def\bZ{\mathbf{Z}}
\def\cB{\mathcal{B}}
\def\cF{\mathcal{F}}

\def\cU{\mathcal{U}}
\def\cL{\mathcal{L}}
\def\cM{\mathcal{M}}
\def\bbeta{\mbox{\boldmath $\beta$}}

\def\bgamma{\mbox{\boldmath $\gamma$}}
\def\bldeta{\mbox{\boldmath $\eta$}}

\def\bkappa{\mbox{\boldmath $\kappa$}}
\def\blambda{\mbox{\boldmath $\lambda$}}
\def\bmu{\mbox{\boldmath $\mu$}}

\def\btheta{\mbox{\boldmath $\theta$}}
\def\brho{\mbox{\boldmath $\rho$}}

\def\bSigma{\mbox{\boldmath $\Sigma$}}
\def\var{\textrm{var}}

\def\log{\textrm{log}}

\def\Poi{\textrm{Poi}}
\def\Unif{\textrm{Unif}}
\def\upp{^{\prime}}

\IfFileExists{upquote.sty}{\usepackage{upquote}}{}

\begin{document}


\titlepage
\title {Estimating Abundance from Counts in Large Data Sets of Irregularly-Spaced Plots using Spatial Basis Functions}
\author{Jay M. Ver Hoef, John K. Jansen\\
\hrulefill \\
$^{1}$ 
National Marine Mammal Laboratory\\
NOAA-NMFS Alaska Fisheries Science Center\\
7600 Sand Point Way NE, Bldg 4\\
Seattle, WA  98115-6349 USA\\
tel: (907) 456-1995 \\
E-mail: jay.verhoef@noaa.gov\\
\hrulefill \\
}

\maketitle


\begin{abstract}

Monitoring plant and animal populations is an important goal for both academic research and management of natural resources. Successful management of populations often depends on obtaining estimates of their mean or total over a region. The basic problem considered in this paper is the estimation of a total from a sample of plots containing count data, but the plot placements are spatially irregular and non randomized.  Our application had counts from thousands of irregularly-spaced aerial photo images.  We used change-of-support methods to model counts in images as a realization of an inhomogeneous Poisson process that used spatial basis functions to model the spatial intensity surface. The method was very fast and took only a few seconds for thousands of images.  The fitted intensity surface was integrated to provide an estimate from all unsampled areas, which is added to the observed counts. The proposed method also provides a finite area correction factor to variance estimation. The intensity surface from an inhomogeneous Poisson process tends to be too smooth for locally clustered points, typical of animal distributions, so we introduce several new overdispersion estimators due to poor performance of the classic one. We used simulated data to examine estimation bias and to investigate several variance estimators with overdispersion.  A real example is given of harbor seal counts from aerial surveys in an Alaskan glacial fjord. 

\noindent \hrulefill 

\end{abstract}


\newpage
\section{Introduction}

Monitoring plant and animal populations is an important goal for both academic research and management of natural resources. Successful management of populations often depends on estimates of their mean or total over a region. Historically, this has been the purview of sampling theory using simple random sampling, stratified random sampling, etc., which are design-based methods.  For design-based methods, sample units are chosen at random, measurements are made or observed from the sample units, and inference is derived from the inclusion probability for sample units (i.e., Horwitz-Thompson estimation).  For overviews, see \citet{Coch:samp:1977} or \citet{Thom:samp:1992}. An alternative approach developed in the early 1960's called geostatistics includes methods such as block kriging \citep{Gand:prob:1959, Gand:opti:1960, Math:Prin:1963}, which also estimates a regional total.  These methods rely on an assumption about a stochastic process that generated the realized observations, and are hence ``model-based.''  Model-based inference relies on estimating parameters for the assumed model, and then forming probability statements (confidence intervals, prediction intervals, etc.) from the fitted model.  In this paper we pursue the model-based approach because samples cannot always be drawn randomly. In particular, we consider counts from aerial photographs, which are difficult and inefficient to randomize, with the basic problem being the estimation of the total count for a region.  

\begin{figure}[H]
	\begin{center}
	\includegraphics[width=450pt]{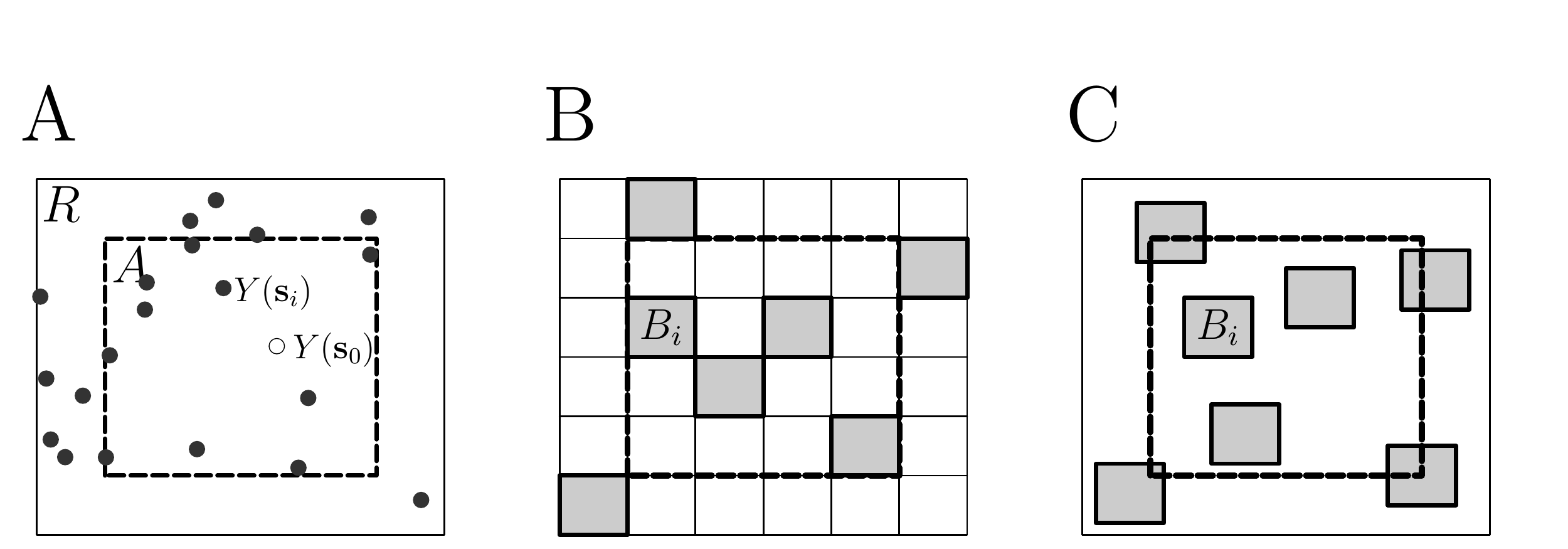}
	\end{center}
	\caption{A. The domain of interest is $R$ (thin solid line) with a spatial random field throughout, where the random variable at location $\mathbf{s}_i$ is denoted $Y(\mathbf{s}_i)$.  Block kriging predicts the average or total for region $A$ (heavy dashed line) B. A finite population version of block kriging. The samples are on a regular grid and a finite number of $\{Y(B_i)\}$
	exhaustively fills both $R$ and $A$, and the goal is to predict $Y(A) \equiv
	\sum_AY(B_i)$ from a sample of $\{Y(B_i)\}\subseteq R$.  C. The situation where
	the sample units have substantial area, but do not form a regular grid within
	$R$ or $A$.\label{FigRegionsSamples}}         
	\end{figure}

The goals and context of this paper are shown in Figure~\ref{FigRegionsSamples}. The situation for block kriging is shown in Figure~\ref{FigRegionsSamples}A.  Let the spatial region of interest be $R$. Any particular location in $R$ is given by $x$- and $y$-coordinates contained in the vector $\bs = [s_x,s_y]^\prime$, and the random variable at the $i$th location is denoted $Y(\bs_i)$. We assume a spatial random field $\{Y(\bs): \bs \in R\}$ \citep[pg. 30]{Cres:stat:1993}. The set $\{Y(\mathbf{s})\}$ is continuous in space and hence infinite.  We use the notation that for vector $\bx=[x_1$, $x_2]^\prime$, $\|\bx\| \equiv \sqrt{x_1^2 + x_2^2}$, so $\|\bs_i - \bs_j \|$ is Euclidean distance between $\bs_i$ and $\bs_j$. If the correlation between $Y(\mathbf{s}_i)$ and $Y(\mathbf{s}_j)$ goes to one as $\|\mathbf{s}_i - \mathbf{s}_j\| \rightarrow 0$, then $\{Y(\mathbf{s}_i)\}$ will form a smooth (differentiable) but random surface.  Suppose that $n$ observed values from the random surface are contained in the vector $\by = [y(\bs_1),\ldots,y(\bs_n)]^{\prime}$ (the solid circles in Figure~\ref{FigRegionsSamples}A). Block kriging uses a linear combination $\blambda^{\prime}\mathbf{y}$ to predict the average or total in block $A$; $Y(A) \equiv \int_A Y(\bu)d\bu$, where this integral is assumed to exist (\citealp[][pg. 23]{Yagl:intr:1962}; \citealp[][pg.106]{Cres:stat:1993}). The salient feature of block kriging is that a model of the autocorrelation of the spatial random field can be estimated from the point level data of the observations, and it is relatively easy to aggregate (through the integral) for an estimator of block $A$. However, extensions to count data have been difficult because data are modeled on a transformed space but the integral is desired on the original space \citep[e.g., see][p. 286]{Cres:stat:1993}.  For example, \citet*{Chri:Waag:baye:2002, Wikl:spat:2002, Mone:Dubr:Bonn:Durb:Guin:Geos:2006} develop maps from count data but do not attempt abundance estimates.                         
 
Counts are often obtained from plots, $B_i$ in Figure~\ref{FigRegionsSamples}B, that have substantial area, are in a regular grid, and exhaustively fill both $R$ and $A$. Here, classical random sampling using design-based inferences are often employed, where a random sample of the observed count on the $i$th block, $y(B_i)$, are used to estimate a total.  To correctly estimate variance, a finite population correction factor is employed, $\hat{\textrm{var}}(\hat{Y}(A)) = (\hat{\sigma}^2/n)(1-f)$ where $\hat{\sigma}^2$ is the sample variance, $f=n/N$ is the fraction of sampled units ($n$) to total sample units ($N$) within $A$, and $1 - f$ is called the finite population correction factor. A model-based, finite-population version of block kriging for the situation shown in Figure~\ref{FigRegionsSamples}B was developed by \citet{Ver:pred:2000, Ver:spat:2008}. In the strict sense, distance between samples is not well-defined because of the non-point nature of each sample $B_i$.  However, models of autocorrelation are often built in this case by using the centroid of each plot.  The main problem considered by \citet{Ver:pred:2000, Ver:spat:2008} was that the traditional formulation of block kriging as shown in Figure~\ref{FigRegionsSamples}A assumed an infinite population.  Hence, if one were to estimate an autocorrelation model for Figure~\ref{FigRegionsSamples}B, and then apply standard block kriging formulas (as has been done in the literature), there is no finite correction factor.  For example, if we sampled all of the plots, then the prediction variance should be zero, but the traditional formulation of block kriging would have nonzero prediction variance. Hence, a finite population version was developed  by \citet{Ver:pred:2000, Ver:spat:2008}, and it performed well and had proper confidence intervals in a variety of situations \citep{Ver:samp:2002}. 
  
Now consider the situation in Figure~\ref{FigRegionsSamples}C.  Here, we would like to use counts from samples with substantial area, $y(B_i)$, to predict at $Y(A)$, but the sample units are not arranged in a regular grid that fills either $R$ or $A$.  Classical random sampling would usually be employed in this situation, with an estimator of the total being $|A|\bar{Y}(B)$, where $\bar{Y}(B)$ is the total count divided by the sampled area, and the variance estimated by $\textrm{var}(\hat{Y}(A)) = (\sigma^2/n)(1-f)$, where here $f=|a|/|A|$, with $|a|$ being the total area sampled. However, what if the samples $Y(B_i)$ are not randomly placed, and may in fact be in some regular pattern that does not form a regular grid that does not exhaustively fill both $D$ and $A$?  The basic problem considered in this paper is the prediction of a total in region $A$ from a sample of $\{Y(B_i)\}$ as in Figure~\ref{FigRegionsSamples}C, where sampling at random is not possible. This is the case for counts from photographs taken from aircraft.  The National Marine Mammal Laboratory of the NOAA-NMFS Alaska Fisheries Science Center developed aerial survey methods to estimate and monitor harbor seal populations in glacial fjords in Alaska. We give more details next as a motivating example.

\subsection{Motivating example}\label{sec:motivate}

\begin{figure}[H]
	\begin{center}
	\includegraphics[width=450pt]{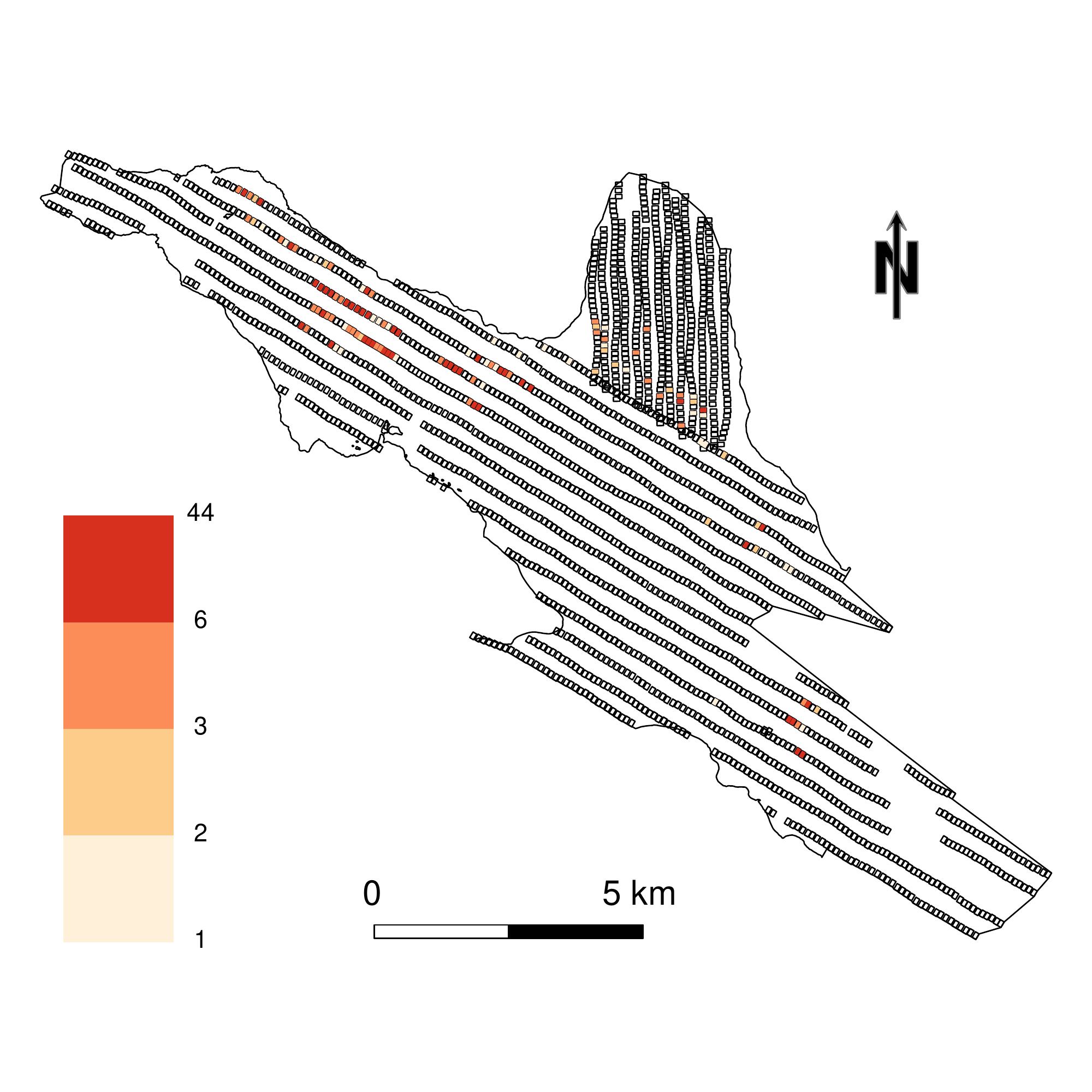}
	\end{center}
	\caption{Example data set of aerial surveys for harbor seals conducted on 11 August 2008 in Icy Bay, Alaska. The outlines of aerial photographs are shown within the study area. Open plots have 0 seals, and darker shaded plots have more seals.  \label{fig:studyArea}}
	\end{figure}

To make the problem concrete, we consider an example that prompted the model development. Aerial surveys were flown over the ice haul-out area of harbor seals in Icy Bay, Alaska.  Ice emanating from tidewater glaciers provides a dynamic expanse of floating ice on which the seals whelp and nurse pups and rest during the molting season.  The aerial platform, a twin-engine Aero Commander Shrike, was flown at 1000 ft and ca. 100 knots on transects with variable spacing that were oriented in two main directions to sample the two main arms of the bay (Figure~\ref{fig:studyArea}), covering about 79 sq km.  A vertically-mounted camera (Nikon D1X with a 60 mm lens) captured an image approximately every 2 seconds through a portal, each covering about 80 $\times$ 120 m at the surface of the water. This firing rate, and the spacing of the transects, allowed for a gap between images of about 30 m end-to-end, and transects varied in spacing from side-to-side, but largely ensured that images were separated from each other; i.e., seals were sampled only once.  The camera was usually turned off when flying over large areas of open water where hauled out seals would necessarily be absent.  This survey was conducted on 20 May, 2004, in the afternoon (1300 to 1430 hr) when seals typically haul out in peak numbers. This is just one data set collected among dozens annually as part of a monitoring program for harbor seals.

Images were georeferenced and embedded as a raster layer in an ArcGIS \citep*{ESRI:ArcG:2009} project allowing individual seals to be spatially marked in a point layer by visually inspecting each image ($n$ = 2080 images).  Footprints showing the extent of each image were generated as polygons (Figure~\ref{fig:studyArea}) in a separate layer and seal points were summed within and assigned to each centroid and exported for statistical analysis.  The spatial extent of each image was assumed to be constant despite small random variation in altitude (max: $\pm$ 30 m) during the survey.  The total spatial extent of each day’s survey effort, over which the intensity surface would be calculated, was delineated by creating a polygon that corresponded to: 1) the coastline of the bay (shorelines from Alaska Department of Natural Resources line shapefiles), 2) an estimate of the location of the face of the glaciers (by connecting points that marked the glacial terminus in each of the northernmost images from every transect, and 3) the extent of the ice field defined as the edge of the images where ice cover (by area) dropped to $<5\%$ by visual estimation.  Areas of open water ($<5\%$ cover) were delineated by “donut-holes” in the overall polygon where the spatial boundaries was defined by the outermost images in which ice cover increased to $ \geq 5\%$.  In other words, to minimize problems with selection bias, any area that could not contain a seal was eliminated by creating the proper boundary.

The 2080 images covered 25.3\% of the study area. Of the 2080 images, 180 of them had nonzero counts, so about 91\% were zero. A total of 1002  seals were observed in the 180 plots. A maximum of 44 seals were counted in a single photograph. The data are summarized spatially in Figure~\ref{fig:studyArea}. 

\subsection{Previous work} \label{sec:prevWork}

Most of the previous work in this area has used Bayesian models.  The literature has concentrated on producing smoothed maps of relative abundance, although going from those smooth maps to an abundance estimate would not seem difficult. In particular, \citet{Wikl:spat:2002} developed Poisson-lognormal models for a continuous surface, but the counts were at a scale that could be considered points, whereas we have counts in plots with substantial areas.  \citet{Thog:Saue:Knut:hier:2004} used a tessellation to create spatial conditional autoregressive (CAR) models \citep{Besa:spat:1974} for neighbors as a spatial random effects model, with the CAR random effects constant within the tessellation, but the model retained point level data for covariates. \citet{Royl:Kery:Gaut:Schm:hier:2007} model on a subsample of a systematic grid, and include detection models, but ultimately use a continuous Poisson-lognormal model for counts. \citet{Barb:Gelf:hier:2007} also use Poisson-lognormal models with known covariates to model the intensity surface.  Note that none of these methods include a finite population correction factor, and while they are all very attractive, we do not adopt any of them for reasons that we describe next.

\subsection{Goals and Organization} \label{sec:GoalsOrg}

Based on this introduction, we desire a total abundance estimator from data like the motivating example that will satisfy several practical conditions. 1) It must be fast to compute, robust, and require few modeling decisions, similar to classical survey methods. Annually, we compute dozens of estimates for data like the example and, depending on the size of the fjord, each may have thousands of photographs. 2) The estimator must use only counts within plots; actual spatial locations of animals are unknown. 3) We are interested in the actual number of seals, not the mean of some assumed process that generated the data. Thus, the estimator must make use of the actual number of seals, and predict to those areas that are unsurveyed. 4) The variance estimator should have a population correction factor that shrinks to zero as the proportion of the study area that gets sampled goes to one.  In our real example, we surveyed approximately 26\% of the study area; in classical sampling, that directly reduces that variance by 26\%. Some fjords that we have surveyed are up to 50\% sampled. 5) The estimator should be approximately unbiased (demonstrated through simulations), and we want valid confidence intervals that cover the true number of seals the correct proportion of times; that is, we use this method dozens of times per year and desire confidence intervals in the frequentist sense. 6) It appears that there is nonstationary variance throughout the area, with large areas of zero counts (no seals).  A variance estimator that accommodates this will be required.  The goals of this manuscript are to develop an estimator to satisfy these criteria.

The rest of this paper is organized as follows.  The estimator is developed from models for spatial point processes and generalized linear mixed models, so we begin with a brief review and then develop the estimator in Section~\ref{SPPEstimator}.  We provide some simulations to validate the estimator and compare variance estimators in Section~\ref{SimExp}. In Section~\ref{sec:Example} we use the estimator on the motivating example of aerial photographs taken of harbor seals in a glacial fjord in Alaska.  We provide some concluding remarks in Section~\ref{sec:DiscConc}.



\section{Model development} \label{SPPEstimator}

From the Introduction and motivating example, note that data are observed at an aggregated support level, but we need a model at the point support level. The reason should be clear; because of the possibility of unbalanced spatial sampling (Figure~\ref{FigRegionsSamples}C), and a real example of it (Figure~\ref{fig:studyArea}), we need to predict and then integrate an abundance density surface continuously throughout the unsampled area. To achieve this, we develop a model-based estimator motivated by an inhomogeneous point process (IPP) that has been integrated to yield a Poisson regression model.  Part of attraction of this framework is that it allows inference on point level support from data on areal support, and then we use the point level support model to make our abundance estimate. We begin with a brief review of the IPP, describe how abundance is related to the intensity surface, and then draw the connection to Poisson regression.


\subsection{Inhomogeneous point process model}

\begin{figure}[H]
	\begin{center}
	\includegraphics[width = .50\maxwidth]{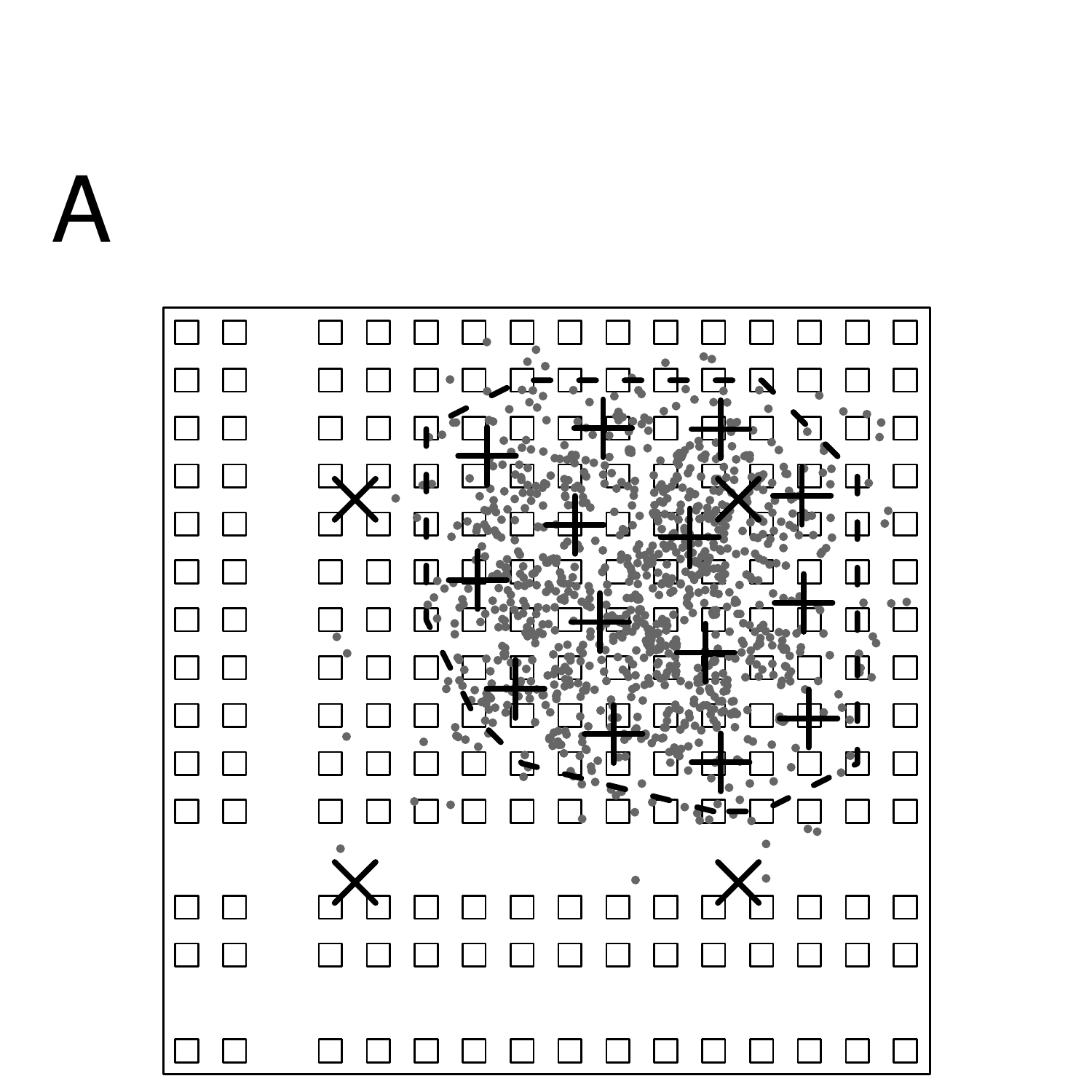} \\
	\includegraphics[width = .50\maxwidth]{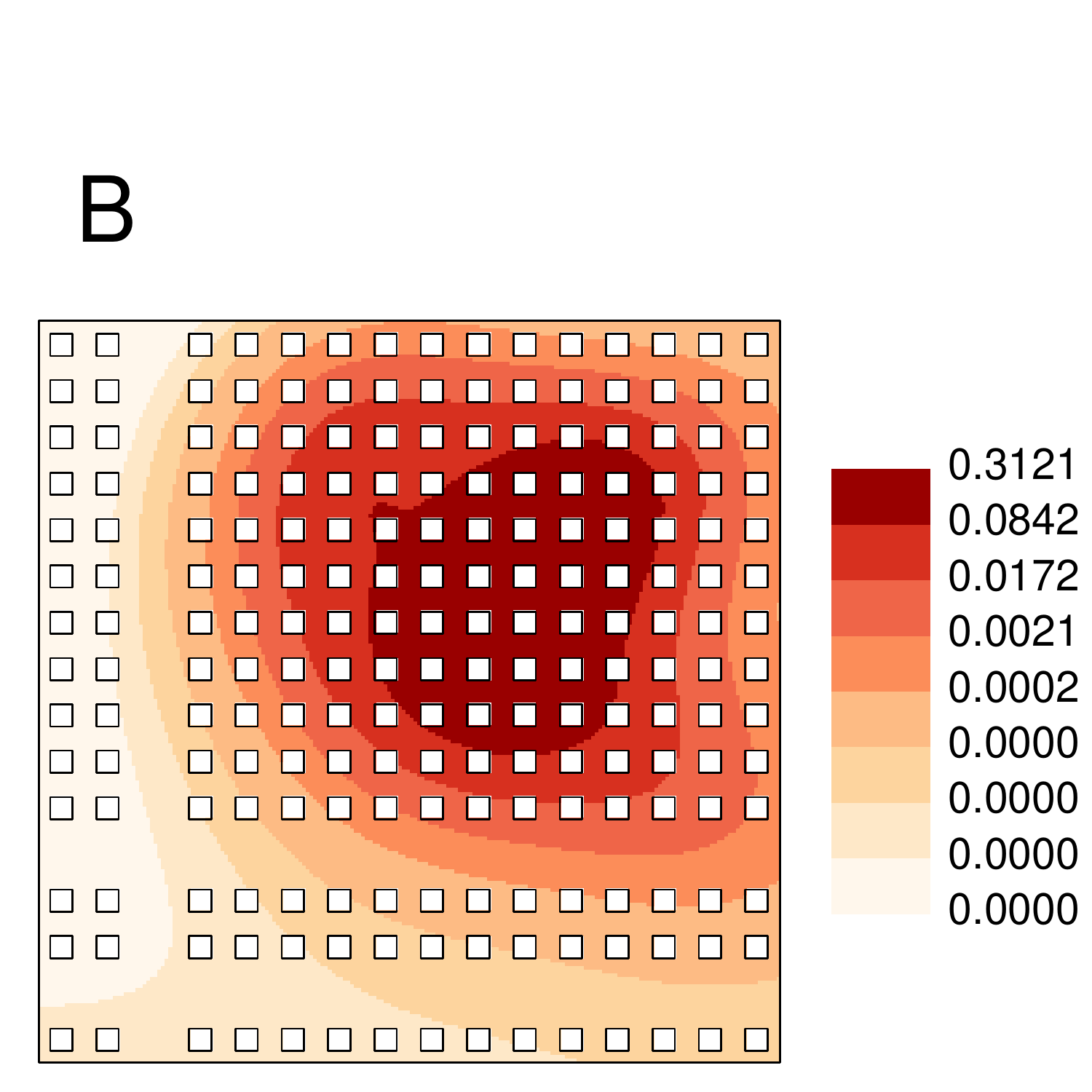}
	\end{center}
  \caption{A. One simulated realization, where all simulated points are shown as grey dots, from simulation experiment 3, described in Section~\ref{sec:Exp3}.  The coarse-scale knot locations are shown with an ``$\times$'' while the fine-scale knots are shown with a ``+''. The fine-scale knots are contained within the convex polygon given by the dashed lines, which bounds the centroids of plots containing nonzero counts. B. The fitted intensity surface throughout $\cU$, scaled to the size of the prediction block, to yield the expected count per prediction block.  \label{fig:SimulationExample}}
	\end{figure}

Assume that locations $\bs=[s_x,s_y]^\prime$ of all individuals in $A$, say $\cS^+ = (\bs_1,\dots,\bs_N)$, are the result of an inhomogeneous Poisson process (IPP) with intensity function $\lambda(\bs|\btheta)$ that varies with $\bs$, where $\btheta$ is a vector of parameters controlling the intensity function. The intensity function is defined as     
\[
\lambda(\bs) = \lim_{|d\bs| \to 0} \frac{E\left( T(d\bs) \right) }{|d\bs|},
\]
where $E(\cdot)$ is expectation, $T(R)$ is the total number of points in planar region $R$, and $|R|$ is the area of $R$. In general, when analyzing IPP data, all of the individuals would be located within $A$, and then inference about $\btheta$ could be made by maximizing the point process log-likelihood \citep*[e.g., ][p. 655]{Cres:stat:1993}. However, this is difficult in our case because $A$ cannot be surveyed in its entirety and individual locations are unknown. For example, a simulated point pattern is shown in Figure~\ref{fig:SimulationExample}, and the plots form a disjointed window on the point pattern that is masked in the areas between the plots, and although we show the point locations within plots, we assume we only have a count for each plot. Ultimately, we will need to estimate the intensity function, but for now we proceed assuming that we have an estimate of the intensity function in the area between plots, as shown in Figure~\ref{fig:SimulationExample}.      

\subsection{Estimating abundance}

The primary quantity of interest is the abundance in a particular block $A \subseteq R$. Because the distribution of individuals is random under the model-based paradigm, there are two types of abundance to consider. First is the {\em expected} abundance in $A$, $\mu(A)=\int_A \lambda(\bu|\btheta)d\bu$, and then there is the {\em realized} abundance $T(A)$ for a given realization of $\cS^+$ from $\lambda(\bs|\btheta)$. Assuming an inhomogeneous point process, $T(A) \sim \Poi(\mu(A))$, which is Poisson distribution with mean $\mu(A)$. An estimate of the {\em expected} abundance is $\hat{\mu}(A) = \int_A \hat{\lambda}(\bu|\btheta) d\bu$, which is often based on plug-in methods from estimates $\hat{\btheta}$ of $\btheta$; i.e.,  $\hat{\mu}(A) = \int_A \lambda(\bu|\hat{\btheta}) d\bu$. 

For an estimate of the realized abundance, consider that the total abundance can be partitioned into observed and unobserved. Assume there are $n$ sample units $B_i \in A$, and let $\cB = \cup_{i = 1}^n B_i$.  Note that some $B_i$ could be outside $A$ but within $R$.  It is also possible that some $B_i$ straddle the boundary of $A$, though we will not consider that problem here. The region within $A$ that was not sampled is $\cU \equiv \overline{\cB} \cap A$, where $\overline{\cB}$ is the complement of $\cB$. Then $T(\cB)$ is the number of observed points and $T(\cU)$ the number of unobserved points and $T(A) = T(\cB) + T(\cU)$. The total $T(A)$ involves predicting $T(\cU) \sim \Poi(\mu(\cU))$, where $\mu(\cU)= \int_{\cU} \lambda(\bu|\btheta) d\bu$. By substituting $\hat{\btheta}$ for $\btheta$, one can use the estimator $\hat{\mu}(\cU)= \int_{\cU} \lambda(\bu|\hat{\btheta}) d\bu$, and, without any observations from $\cU$, we use the mean as a predictor $\widehat{T}(\cU) = \hat{\mu}(\cU)$. Hence an estimator of the total is
\begin{equation}\label{eq:AbunEst}
	\widehat{T}(A) = T(\cB) + \widehat{T}(\cU),
\end{equation}
for making inference to the realized abundance $T(A)$. Note that, as $T(\cB) \rightarrow T(A)$, then $\widehat{T}(A) \rightarrow T(A)$, so an estimator of this form satisfies condition 3 in Section~\ref{sec:GoalsOrg}. Making such inferences involves first estimating the intensity function $\lambda(\bs|\btheta)$, and also incorporating the uncertainty of estimating $\lambda(\bs|\btheta)$.  So our immediate goal is to infer $\lambda(\bs|\btheta)$ from data on areal support, which we describe in the next sections.

\subsection{From IPP to Poisson Regression}\label{sec:IPPtoPoi}

Suppose that we have a smooth spatial surface $\lambda(\bs|\btheta)$ that varies with spatial location $\bs$ and is controlled by parameters $\btheta$; this is the intensity surface.  This surface may be integrated over some compact region, such as the plot $B_i$, and this forms the mean of an IPP.  Let $Y(B_i)$ be a random variable for a count in $B_i$, then $Y(B_i) \sim \Poi(\mu(B_i))$, where
\begin{equation}\label{eq:muBi}
	\mu(B_i) = \int_{B_i}\lambda(\bu|\btheta) d\bu.
\end{equation}
Now, let $\bs_i$ be the centroid of plot $B_i$. If the area of $B_i$ is small compared to the survey area $A$, and if $\lambda(\bu|\btheta)$ is smooth (i.e., changing slowly within $B_i$), then \citet{Berm:Turn:appo:1992} show that a reasonable approximation for (\ref{eq:muBi}) is,
\begin{equation}\label{eq:muBiPoint}
	\mu(B_i) = |{B_i}|\lambda(\bs_i|\btheta),
\end{equation}
where $|{B_i}|$ is the area of $B_i$. This is an important assumption and is part of the general problem of change-of-support; see \citet{Gotw:Youn:comb:2002}, \citet*[][Chapter 6]{Bane:Carl:Gelf:hier:2004} and \citet{Wikl:Berl:comb:2005}.  For example, \citet{Bril:spat:1990, Bril:exan:1994} shows an early attempt at creating a continuous surface from count data in census tracts.  

The mean of $Y(B_i)$ can then be modeled with a log link function, forming a GLM with offset $\log(|B_i|)$,
\[
	\log(\mu(B_i)) = \log(|B_i|) + \log(\lambda(\bs_i|\btheta)).
\]
Now we use spatial radial-basis functions to model $\lambda(\bs_i|\btheta)$. Let $\bs_i$ be the centroid of plot $B_i$.  Then 
\begin{equation}\label{eq:logLambda}
	\log(\lambda(\bs_i|\btheta)) = \beta_0^* + \bz(\bs_i)^\prime\bgamma,
\end{equation}
where $\bz(\bs_i)$ is a vector of covariates at location $\bs_i$ and $\bgamma$ is a parameter vector of fixed effects. The spatial basis functions will form the values of $\bz(\bs_i)$.

There has been increasing interest lately in spatial models that use radial basis functions. Suppose there is a set of fixed points in the study area, $\{\bkappa_j;j=1,\ldots,K\}$, called ``knots.'' Let $\bz(\bs_i)\upp$ be a row vector where the $j$th item contains a radial basis function value $C(\| \bs_i-\bkappa_j)\|;\rho)$. For example, we will use $C(h;\rho)=\exp(-h^2/\rho); \rho > 0$, which is a Gaussian basis function. A flexible surface is created by taking a linear combination of the radial basis functions. The surface value at location $\bs_i$ depends on parameters $\bgamma$ and $\rho$ as $\bz_\rho(\bs_i)\upp\bgamma$, and we attach the subscript to show that values in $\bz$ depend on $\rho$.  Using radial basis functions can be viewed as a semiparametric approach to spatial modeling \citep*{Rupp:Wand:Carr:semi:2003}, and they have been used for models with non-Euclidean distance measurements \citep*[see, e.g.,][]{Wang:Rana:low:2007} and for computational efficiency for large data sets \citep*{Cres:Gard:fixe:2008}. 

To make the model more flexible, following \citet{Cres:Gard:fixe:2008}, we considered radial basis functions at two scales.  Let the ``coarse'' scale knots be $\{\bkappa_{C,j};j=1,\ldots,K_C\}$.  Let the fine scale knots be $\{\bkappa_{F,j};j=1,\ldots,K_F\}$, where generally $K_F \geq 4K_C$. Note that \citet{Cres:Gard:fixe:2008} use 3 scales with approximately 3 times as many knots at the next finer scale. Here, because we only have two scales, we use 4 times as many knots at the finer scale.  The knots are generally spread out more or less regularly throughout the study area; more details on an algorithm for knot locations are given in Section~\ref{sec:Knots}.

Consider the log-linear model
\begin{equation}\label{eq:mixmod}
 \log(\blambda) = \bX\btheta = \bW\bbeta + \bZ_C\bgamma_C + \bZ_F\bgamma_F,
\end{equation}
where $\bX = [\bW|\bZ_C|\bZ_F]$, $\btheta = [\bbeta^{\prime},\bgamma_C^{\prime},\bgamma_F^{\prime}]^{\prime}$, $\bZ = [\bZ_C|\bZ_F]$, $\bgamma = [\bgamma_C|\bgamma_F]$ and the $j$th column of $\bZ_C$ has $C(\| \bs_i-\bkappa_{C,j})\|;\rho_C)$ as the $i$th element,  and the $j$th column of $\bZ_F$ has $C(\| \bs_i-\bkappa_{F,j})\|;\rho_F)$ as the $i$th element. We will not consider any covariates in our model, allowing all spatial variation to be modeled through the spatial basis functions, although future development could easily accommodate covariates here. From (\ref{eq:logLambda}), we only consider an overall constant, $\bx(\bs_i)^\prime\bbeta = \beta_0$.  Also, we assume all plots are the same size, $|{B_i}| = |B|$.  Then we can  write,
\begin{equation}\label{eq:logMuBiPoint}
	\log(\mu(B_i)) = \beta_o + \log(|B|) + \bz(\bs_i)^\prime\bgamma,
\end{equation}
where $\bz(\bs_i)$ is the $i$th row of $\bZ$. Model (\ref{eq:logMuBiPoint}) will form the basis for estimation and prediction throughout the rest of this paper.

\subsection{Knot Selection}\label{sec:Knots}

To place coarse scale knots, a systematic grid of points was generated within $A$, and K-means clustering \citep{MacQ:some:1967} on the coordinates was used to create $K_C$ groups.  Because K-means clustering minimizes within-group variance while maximizing among-group variance, the centroid of each group tends to be regularly spaced; i.e., it is a space-filling design that can work well when the region $A$ has an irregular boundary, as in our example data set (Section~\ref{sec:motivate}).  We also used K-means clustering placing $K_F$ fine scale knots, but the systematic grid was generated within a minimum convex polygon that contained all non-zero counts intersected with $A$; this polygon was defined on the centroids of plots with nonzero counts, and an example can be seen in Figure~\ref{fig:SimulationExample}.  We found that this helped ensure convergence of the algorithm.  If there are too many basis functions with a small range centered in a large area that is all zeros, the fitting algorithm that we describe next would fail to converge.  The effect of knot numbers, both $K_C$ and $K_F$, are examined in the simulation experiments in Section~\ref{SimExp}.  Other methods for spatial knot placement could be used; for example see \citet{Nych:Salt:desi:1998}. The software PROC GLIMMIX \citep{SAS:stat:2008} generates spatial knots using vertexes of a k-d tree \citep{Frie:Bent:Fink:algo:1977}. Regarding the number of spatial knots, \citet[][pg. 255]{Rupp:Wand:Carr:semi:2003} recommend $K_C + K_F = n/4$, with no less than 20 and no more than 150.

\subsection{Parameter Estimation}\label{sec:Est}

Recall that the $i$th plot $B_i$ is very small in relation to $A$, and we let $\bs_i$ be the centroid of the $i$th plot. The count in the $i$th plot is random, denoted $Y(B_i)$, and starting from an inhomogeneous Poisson process, from (\ref{eq:logMuBiPoint}) we assume that $Y(B_i)$ has a Poisson distribution with mean $\mu(B_i) =  \exp(\beta_o + \bz(\bs_i)^\prime\bgamma)$. This is Poisson regression, more generally formed as a generalized linear model (GLM) \citep{McCu:Neld:gene:1989},
\begin{equation} \label{glm.mean}
	 E(\bY|\btheta) = g^{-1}(\bX\btheta) = g^{-1}(\bldeta) = \bmu,
\end{equation}
where $\bY = (Y(B_1),\dots,Y(B_n))$, and $\bX$ and $\btheta$ were defined following (\ref{eq:mixmod}). Conditional on fixed $\brho$ values contained in the $\bZ$ part of $\bX$, iteratively weighted least squares (IWLS) \citep{Neld:Wedd:gene:1972} provides maximum likelihood estimation for $\btheta$.  Recall that the negative log likelihood for Poisson regression is
\begin{equation} \label{eq:PoiLL}
  \ell(\brho,\btheta;\by) = \sum_{i=1}^{n} |B_i|\exp(\bx_{\brho}(\bs_i)\upp\btheta) - y_i\log|B_i| - y_i\bx_{\brho}(\bs_i)\upp\btheta,
\end{equation}
where $\bx_{\brho}(\bs_i)$ is the $i$th row of $\bX$ in (\ref{eq:mixmod}) with the specific case being (\ref{eq:logMuBiPoint}).  Here, we show the dependence of that row on $\brho$ values. An iterative algorithm using block-wise coordinate descent for minimizing the negative likelihood is,
\begin{itemize}
	\item condition on $\brho = [\rho_C,\rho_F]^\prime$ and use IWLS to estimate $\btheta$,
  \item embed the IWLS estimation in a numerical optimization of (\ref{eq:PoiLL}) for $\brho$.
\end{itemize}
This optimization routine over just two parameters, $\brho$, converges quickly and can use existing Poisson regression software for the IWLS update, so it satisfies the speed requirement of condition 1 in Section~\ref{sec:GoalsOrg}.  To help ensure convergence, we constrained $\rho_F$ to be between 0.5 and 3 times the minimum distance between any two knots in $\{\bkappa_F \}$, and constrained $\rho_C$ to be greater than $\rho_F$ but less than 3 times the minimum distance between any two knots in $\{\bkappa_C \}$.  Optimization used the glm() and optim() functions in R \citep{R:Deve:Core:ALan:2014}, where optim() used the Nelder-Mead optimization algorithm \citep{Neld:Mead:simp:1965}.  To ensure boundary conditions, say $a$ as a lower bound and $b$ as an upper bound for one of the elements in $\brho$, we used a transformation $\rho = a + (b-a)\exp(\rho^*)/(1 + \exp(\rho^*))$, and then optimized for unconstrained $\rho^*$ (note that $a$ was a sliding lower boundary for $\rho_C$, but it would stabilize as $\rho_F$ found its optimum).

Also, note the connection to the Janossy density for IPP \citep[see, e.g., ][p. 655]{Cres:stat:1993}.  For some area $B$ with $Y \in 1, 2, \dots$ points at locations $\{\bs_k; k = 1, 2, \ldots, Y\}$ within $B$, the Janossy likelihood is,
\begin{equation} \label{eq:Janossy}
  \cL(\btheta;B) = \left\{ \prod_{k=1}^Y \lambda(\bs_k|\btheta,\brho) \right\} 
    \exp \left\{ - \int_B \lambda(\bu|\btheta,\brho)d\bu \right\}.
\end{equation}
From Section~\ref{sec:IPPtoPoi}, we are assuming that the plots are small enough so that the intensity function is approximately constant within plot, with the intensity value taken from the intensity surface at the centroid of the plot.  Using this approximation, then from (\ref{eq:Janossy}) the negative loglikelihood for all plots is
\[
 \ell(\brho,\btheta;\by) \approx \sum_{i=1}^n - y_i \log[\lambda(\bs_i|\btheta,\brho)] + |B_i|\lambda(\bs_i|\btheta,\brho), 
\]
and, when using model (\ref{eq:logMuBiPoint}) for $\lambda(\bs_i|\btheta,\brho)$, this makes it apparent that minimizing (\ref{eq:PoiLL}) for $\btheta$ and $\brho$ is an approximation to maximizing (\ref{eq:Janossy}).  This connection is important because, in Section~\ref{sec:VarEst}, we use results from maximum likelihood estimation of the Janossy density in IPP literature to obtain variance estimates.  Note that other approaches may be taken, including penalized splines \citep{Rupp:Wand:Carr:semi:2003} or Bayesian approaches (see Section~\ref{sec:prevWork}).

\subsection{Plug-in Abundance Estimator}

Denote $\hat{\btheta}$ and $\hat{\brho}$ as the maximum likelihood estimates from Section~\ref{sec:Est}. Going back to our estimator, recall that $\widehat{T}(A) = T(\cB) + \widehat{T}(\cU)$, and we will use our parameter estimates from Section~\ref{sec:Est} to obtain the predictor $\widehat{T}(\cU) =\mu(\cU)=\int_\cU \lambda(\bu|\hat{\brho},\hat{\btheta})d\bu$, where $\lambda(\bu|\hat{\brho},\hat{\btheta}) = \exp(\bx_{\hat{\brho}}(\bu)\upp\hat{\btheta})$. The integral can be approximated with a dense grid of $n_p$ points within $\bu_j \in \cU$,
\begin{equation} \label{eq:ThatA}
  \widehat{T}(A) = T(\cB) + \sum_{j=1}^{n_p}|U_i|\exp(\bx_{\hat{\brho}}(\bu_j)\upp\hat{\btheta}),
\end{equation}
where $|U_i|$ is a small area around each $\bu_j$.  We generally assume all $|U_i|$ are equal to $|\cU|/n_p$, yielding a 2-dimensional Riemann integral approximation, which is sufficient if $n_p$ is large.  Better approaches using numerical integration by quadrature could also be used.

\subsection{Variance Estimation} \label{sec:VarEst}

The mean-squared prediction error of (\ref{eq:AbunEst}) is
\begin{equation} \label{eq:varT}
		\cM(\hat{T}(A)) = E[(\hat{T}(A) - T(A))^2] = E[(\hat{T}(\cU) - T(\cU))^2]
\end{equation}
Note that as $\cU \cap A \to \varnothing$, then we count all animals in $A$, and $\cM(\widehat{T}(A)) \to 0$, so that this estimator satisfies condition 4 in Section~\ref{sec:GoalsOrg}. Thus, a finite population correction factor is automatically embedded in the variance estimator.  Also, $\hat{T}(\cU)$ depends on random counts in $\cB$, while $T(\cU)$ depends on random counts in $\cU$.  Under the IPP assumption, these will be independent from each other. Further, assume that $\hat{T}(\cU)$ is an unbiased predictor, so $E[\hat{T}(\cU)] = E[T(\cU)]$. Then $\cM(\hat{T}(A)) = E[(\hat{T}(\cU)^2] - 2E[T(\cU)]^2 + E[T(\cU)^2]$, or 
\[
		\cM(\hat{T}(A)) = \var[T(\cU)] + \var[\hat{T}(\cU)].
\]
For the IPP, $\var[T(\cU)] = \mu(\cU)$, and this is estimated with
\begin{equation} \label{eq:HatMuCU}
  \hat{\mu}(\cU) = \frac{|\cU|}{n_p}\sum_{i=1}^{n_p} \exp[\bx_{\hat{\brho}}(\bs_i)^\prime\hat{\btheta}]
\end{equation} 
over the same fine grid of points used in (\ref{eq:ThatA}). Recall that $\hat{T}(\cU) = \int_{\cU}\exp[\bx_{\hat{\brho}}(\bu)^\prime\hat{\btheta}] d\bu$. Define a vector $\bc$ where the $i$th element of $\bc$ is
\[
	\frac{\partial \hat{T}(\cU)}{\partial \theta_i} = \int_{\cU}x_i(\bu)\exp[\bx_{\hat{\brho}}(\bu)^\prime\hat{\btheta}] d\bu.
\]
We approximate this integral with
\begin{equation} \label{eq:ci}
	\frac{\partial \hat{T}(\cU)}{\partial \theta_i} \approx 
    \frac{|\cU|}{n_p}\sum_{i=1}^{n_p} x_i(\bs_i)\exp[\bx_{\hat{\brho}}(\bs_i)^\prime\hat{\btheta}],
\end{equation}
where the sum is over a dense grid of $n_p$ prediction points in the unsampled area. Using the delta method \citep{Dorf:a:1938, Ver:who:2012}, $\var[\hat{T}(\cU)] = \bc^\prime \bSigma \bc$,  where $\bSigma = \var(\hat{\btheta})$.  A similar result is given by \citet{John:Laak:Ver:mode:2010} in a distance sampling context.  Then, as shown by \citet{Rath:Cres:asym:1994}, if $\hat{\btheta}$ is a maximum likelihood estimator from the Janossy density for the IPP, then an estimator of $\bSigma$ is
\[
  \hat{\bSigma}= \left[\sum_{i=1}^n\int_{B_i} \bx_{\hat{\brho}}(\bu)\bx_{\hat{\brho}}({\bu})^\prime
    \exp[\bx_{\hat{\brho}}(\bu)^\prime \hat{\btheta}] d\bu\right]^{-1}.
\]
Assuming that $|B_i| = |B| \ \forall \ i$ is small, this can be approximated as
\begin{equation} \label{eq:SigmaHat}
  \hat{\bSigma} =  \left[ |B| \sum_{i=1}^{n} \bx_{\hat{\brho}}(\bs_i)\bx_{\hat{\brho}}({\bs_i})^\prime
    \exp(\bx_{\hat{\brho}}(\bs_i)^\prime \hat{\btheta}) \right]^{-1}.
\end{equation}
Note that, in (\ref{eq:SigmaHat}), variances may become large if the dimension of $\btheta$ is too high (due to overfitting from too many knots). Through simulations, we will investigate the following variance estimator,
\begin{equation} \label{eq:nu1}
  \hat{\cM}(\hat{T}(A)) = \hat{\mu}(\cU) + \bc^\prime\hat{\bSigma}\bc.
\end{equation}
where $\hat{\mu}(\cU)$ is given by (\ref{eq:HatMuCU}), elements of $\bc$ are given by (\ref{eq:ci}), and $\hat{\bSigma}$ is given by (\ref{eq:SigmaHat}).  Equation (\ref{eq:nu1}) has a nice interpretation by decomposing the variance into the prediction of the total due to fixed intensity surface $\hat{\mu}(\cU)$ (given the regression parameters $\btheta$), plus the variance in estimating the regression parameters $\btheta$.  Note that we have not taken into account the estimation of $\brho$.  While this would be desirable, we use $\hat{\brho}$ as plug-in estimators for now.  This is similar to geostatistical models where covariance parameters are first estimated from the data, and then used for subsequent prediction \citep[see, e.g.,][p. 263]{Scha:Gotw:stat:2005}.  While this is not ideal, and can be the subject of further research, our simulations show that it has little consequence for the type of data that we analyze.


\subsection{Overdispersion} \label{sec:overdispersion}

Animals (as well as other spatially patterned points) are often clustered at very fine spatial scales. For animals, this might occur as mother-offspring pairs, clustering around locally desirable habitats, etc.  The inhomogeneous intensity surface estimated in the foregoing discussion will be unlikely to capture this fine scale clustering, which will contribute to the overall variance, and without considering it, the confidence intervals on abundance estimates will be too short. Various estimators of overdispersion for count models have been proposed, and the negative binomial and quasi-Poisson are commonly used \citep[e.g.,][]{Ver:Bove:quas:2007}; see \citet{Hind:Deme:over:1998} for an overview.  Here, we consider quasi-type models, where, if the mean is $\phi$, then the overdispersion is constant multiplier, $\omega$, so the variance is $\omega \phi$.  As we demonstrate next, some form of robust estimation or further modeling is required because overdispersion changes through space.  In the negative binomial context, robust but nonspatial estimation of can be found in \citet{Moor:Tsia:robu:1991}, and nonparametric estimation is found in \citet{Gijb:Pros:Clae:nonp:2010}. Our situation is different than general robustness because we want to either trim residuals based on data with low expected values, or downweight them.  We describe several estimators next, and compare them in simulations.

Let $\phi_i = E(\bY_i|\hat{\bbeta}) = g^{-1}(\bx_i\hat{\bbeta}) = \exp(\bx_i\hat{\bbeta})$ be the fitted intensity surface value for the $i$th plot, where $\bx_i$ is the $i$th row of $\bX$.  Denoting $y_i$ as the observed value for the $i$th plot, we considered four different ways to estimate overdispersion:

\begin{itemize}

\item The traditional estimator:
\[
\omega_{OD} = \max\left(1, \frac{1}{n-q}\sum_{i=1}^n\frac{(y_i-\phi_i)^2}{\phi_i}\right),
\]
where $q$ is the rank of $\bX$.
\item A linear regression estimator.  Under the Poisson model, the variance is equal to the mean.  By regressing the squared residuals against the fitted value, any slope greater than one would be evidence of overdispersion.  The linear regression is set up with a zero intercept, so the model is $(y_i - \phi_i)^2 = \omega\phi_i$.  We used weighted least squares to obtain the estimator,
\[
\omega_{WR} = \max\left(1, \arg\underset{\omega}{\min}\sum_{i=1}^n\sqrt{\phi_i}[(y_i - \phi_i)^2 - \omega\phi_i]^2\right),
\]
where $\sqrt{\phi_i}$ were the weights.  Notice that generally, this may not be a desirable estimator.  Values with small expectations have virtually no effect on the slope, whereas values with larger expectations will have a great deal of leverage.  In our case, this is a desirable feature, as discussed earlier.  In fact, we create additional weight for values with large expectation by using $\sqrt{\phi_i}$.
\item Estimator based on a trimmed mean of squared Pearson residuals from the upper quantile of fitted values.  Let $\cF = \{\phi_1, \phi_2, \ldots, \phi_n\}$ be an unordered set of expected values for the $n$ observed counts, and $\{\phi_{(1)}, \phi_{(2)}, \ldots, \phi_{(n)}\}$ be the set of ordered values, from smallest to largest, where $\phi_{(1)}=\min(\cF)$ and $\phi_{(n)}=\max(\cF)$. Also, if $\phi_{(i)} = \phi_j$, then $y_{(i)} = y_j$; that is, the observed values are ordered by their fitted values as well. Let $0 \le p < 1$ be some proportion, then
\[
\omega_{TG}(p) = \max\left(1, \frac{1}{n - \lfloor np \rfloor}
	\sum_{i = \lfloor np \rfloor + 1}^n\frac{(y_{(i)}-\phi_{(i)})^2}{\phi_{(i)}}\right),
\]
where $\lfloor x \rfloor$ rounds $x$ down to the nearest integer.  That is, the proportion $p$ of the squared Pearson residuals with the lowest fitted values are trimmed from the overdisperson computation.
\end{itemize}

\begin{figure}[H]
		\begin{center}
		\includegraphics[width = .40\maxwidth]{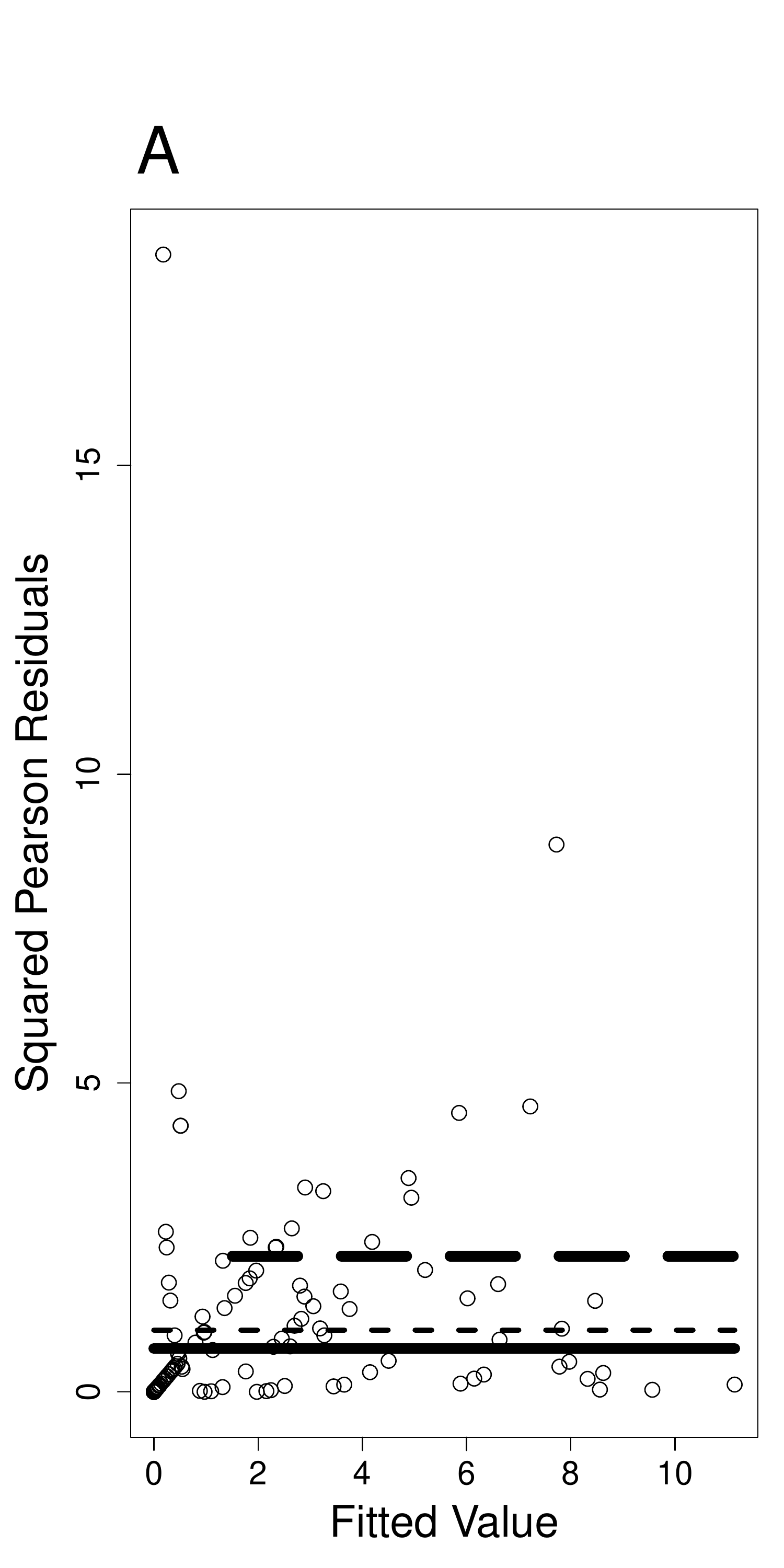}
		\includegraphics[width = .40\maxwidth]{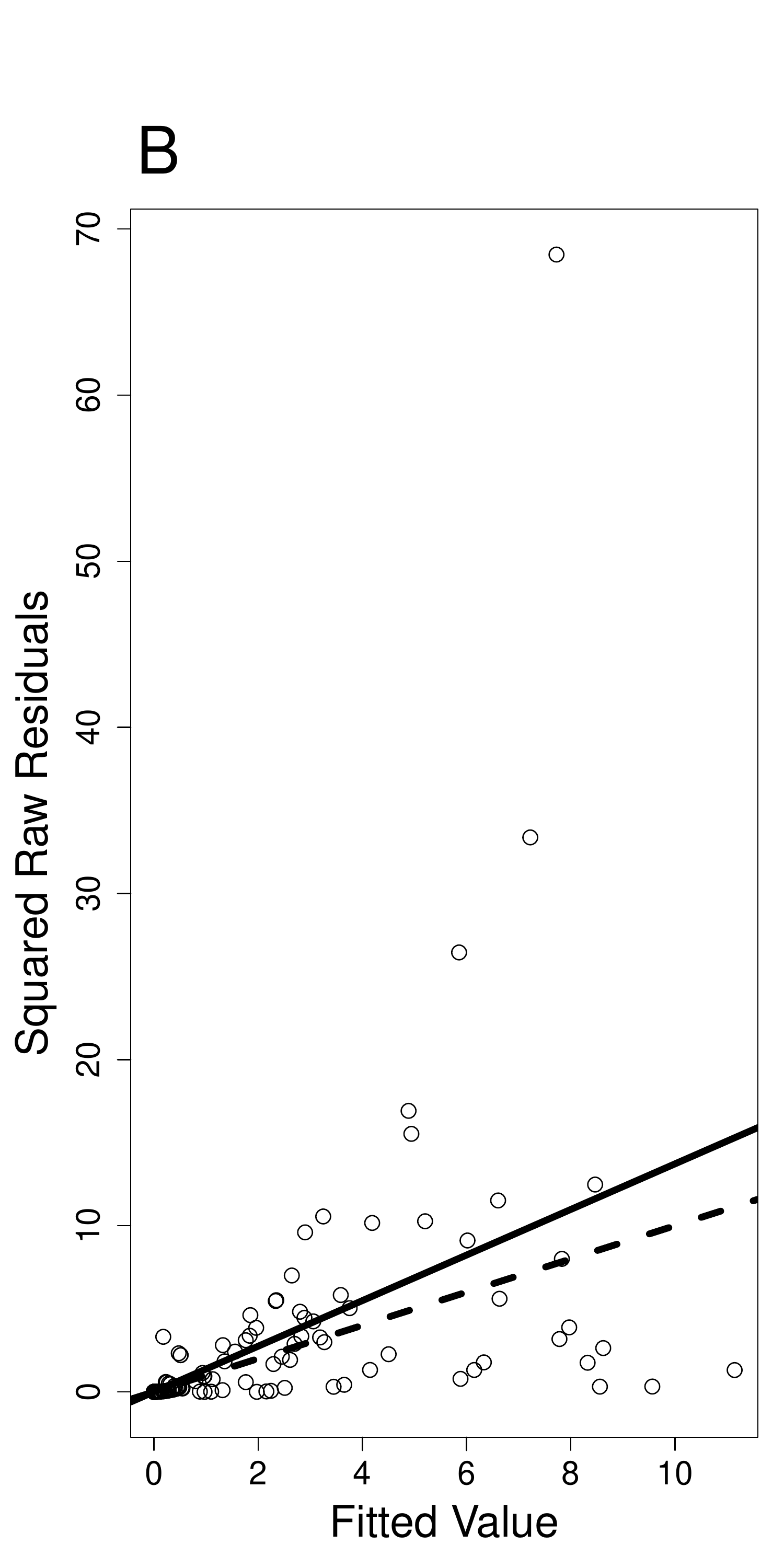}
		\end{center}
		\caption{A. Squared Pearson residuals $(y_i-\phi_i)^2/\phi_i$ plotted against the fitted values $\phi_i$ for the example simulation in Figure~\ref{fig:SimulationExample}. The short-dashed line is constant at 1, and the traditional overdispersion estimator is the solid line below the dashed line.  The upper long-dashed line is the constant value of the trimmed overdispersion estimator, where only the upper 25\% of the ordered values of the fits were used, and the line starts at the lowest of these fitted values.  B. Squared raw residuals $(y_i-\phi_i)^2$ plotted against the fitted values $\phi_i$. The regression estimator of overdispersion is the slope of the solid line, and the dashed line is the one-to-one line.  \label{fig:residDisp}}
		\end{figure}

Examples of the overdisperson estimators are shown in Figure~\ref{fig:residDisp}, which were taken from the data seen in Figure~\ref{fig:SimulationExample}.  The traditional estimator can be viewed as a constant fit (the average value) through all of the squared Pearson residuals for all fitted values, so this is shown as a horizontal solid line, the one that is below the short-dashed line (whose value is constant at one) in Figure~\ref{fig:residDisp}A.  Note especially the wide divergence in squared Pearson residuals for low expected values.  This is not surprising because we are dividing by very small numbers, so any count greater than zero will have a very large residual.  This instability, along with the fact that these values do not really contribute much to overall abundance, leads to estimator $\omega_{TG}(p)$.  Here, we trim off the lowest expected values.  Trimming off the lowest 75\%, and averaging the rest, can be viewed as a constant fit (horizontal line) through the squared Pearson residuals for the upper 25\% of fitted values, and is shown as the long-dashed horizontal line that is above the short-dashed line in Figure~\ref{fig:residDisp}A.  The other idea is to treat raw squared residuals, $(y_i - \phi_i)^2$ as a response variable in a zero-intercept regression, where the predictor variable is the fitted value $\phi_i$.  This is shown as the solid line in Figure~\ref{fig:residDisp}B, which is above the one-to-one line.  The estimated slope of this line is taken as the overdispersion estimate.  Similar to trimming in $\omega_{TG}(p)$, the regression estimator $\omega_{WR}(p)$ downweights residuals with small expected values by forcing the line through zero, and it eliminates division by very small numbers. In fact, we considered weighted regression to add even more weight to higher fitted values.  After some trail and error, we used weights $\sqrt{\phi_i}$, but this is clearly an area for further research.

With these three overdisperson estimators, we have several variance estimators of the abundance estimator (\ref{eq:AbunEst}) at our disposal, 
\begin{equation} \label{eq:vkk}
  \widehat{\var}(\hat{T}(A))_k \equiv \omega_k\hat{\cM}(\hat{T}(A)), 
\end{equation}
where $k$ = $OD$, $WR$ or, $TG$, and $\omega_{TG}$ has the additional trimming parameter $p$. We include one more estimator using the same logic applied to the IPP variance estimator as the trimmed overdispersion estimator $\omega_{TG}(p)$.  If $\phi_{\lfloor np \rfloor}$ represents the smallest fitted value summed in $\omega_{TG}$, then we computed (\ref{eq:SigmaHat}) using only those $i$ sites whose values satisfied $\exp(\bx_{\hat{\brho}}(\bs_i)^\prime\hat{\btheta}) >\phi_{\lfloor np \rfloor}$. Let us call this $\widetilde{\bSigma}$, which when substituted into (\ref{eq:nu1}) and combined with $\omega_{TG}$ yields
\begin{equation} \label{eq:vTL}
  \widehat{\var}(\hat{T}(A))_{TL} \equiv \omega_{TG}(\hat{\mu}(\cU) + \bc^\prime\widetilde{\bSigma}\bc). 
\end{equation}

For confidence intervals, note that from (\ref{eq:ThatA}), the estimate is a sum of a large number of lognormal variates.  That is, we can assume that $\hat{\btheta}$ are normal because they are maximum likelihood estimates.  If each summand in (\ref{eq:ThatA}) was independent, then (\ref{eq:ThatA}) would converge to normality because of the central limit theorem, but due to correlation, the distribution is unknown and may be asymmetric.  We investigated this by simulating (\ref{eq:ThatA}) using $\hat{\bSigma}$ from (\ref{eq:SigmaHat}) as estimated from various data sets.  In all cases, (\ref{eq:ThatA}) was skewed, and a log transformation made the distribution approximately normal. Thus, we recommend computing confidence intervals on the log scale, and then back-transforming.  Using the delta method \citep{Dorf:a:1938, Ver:who:2012}, an approximate 100$(1-\alpha)$\% level confidence interval is
\begin{equation} \label{eq:CI}
  \exp \left( \log(\hat{T}(A)) \pm \frac{z_{\alpha/2}\sqrt{\widehat{\var}(\hat{T}(A))_k}}{\hat{T}(A)} \right),
\end{equation}
for $k$ = $OD$, $WR$, $TG$ or $TL$, where $z_{\alpha/2}$ is the upper $\alpha/2$ percentage point of a standard normal distribution.  Note that this also yields the desirable property that the lower bound of the confidence interval is always greater than 0.


\section{Simulation experiments}\label{SimExp}

We simulated data under four different conditions to examine the performance of the abundance estimator and the variance estimators of abundance.  In all experiments, data were simulated in the region $A = \{(x,y): x \in [0,10] \times y \in [0,10]\}$. For each experiment, data sets were simulated 1000 times, with the number of points and their spatial locations changing, but the sample units were held fixed as shown in Figure~\ref{fig:SimulationPlots}.


\subsection{Evaluating the experiments} \label{sec:EvalExp}

For each experiment described below, the expected number of simulated points was near 1000.  Let $T_t$ be the actual number of points simulated in the $t$th simulation, let $\hat{T_t}$ be the estimator of the total from (\ref{eq:ThatA}), and let $\hat{v}_{k,t}$ be a variance estimator given in (\ref{eq:vkk}) or (\ref{eq:vTL}) for $k$ = $OD$, $WR$, $TG$ or $TL$,  for the $t$th simulation.  The performance of the abundance estimator was evaluated in three ways:

\begin{itemize}

\item Bias for an experiment was computed as,
\[
\frac{1}{1000}\sum_{t=1}^{1000} \hat{T_t} - T_t.
\]

\item Root-mean-squared prediction error (RMSPE) was computed as,
\[
\sqrt{\frac{1}{1000}\sum_{t=1}^{1000} (\hat{T_t} - T_t)^2}.
\]

\item Coverage of 90\% confidence interval (CI90) was computed as,
\[
\frac{1}{1000}\sum_{t=1}^{1000} I\left(\exp(\log(\hat{T_t}) - 1.645\sqrt{\hat{v}_{k,t}}/\hat{T}_t) < T_t 
\; \& \;  T_t < \exp(\log(\hat{T_t}) + 1.645\sqrt{\hat{v}_{k,t}}/\hat{T}_t) \right),
\]
where $I(\cdot)$ is the indicator function, equal to one if the argument is true, otherwise it is zero. Note that (CI90) can be computed for each $k$ in (\ref{eq:vkk}) and (\ref{eq:vTL}), which we denote as CI90$_k$ in the tables that summarize the experiments.

\end{itemize}

For all experiments we used $\omega_{TG}(0.75)$, but investigate the effect of $p$ in Experiment 4. We also investigated knot density by changing $K_C$ and $K_F$, but always using the algorithm described in Section~\ref{sec:Knots}. For each experiment, we included bias, RMSPE, and CI90 for simple random sampling (SRS), as described in the Introduction.  We realize that SRS is inappropriate for these data, but it provides a convenient benchmark for comparison.


\subsection{Experiment 1: Inhomogeneous spatial point process with regular sampling} \label{sec:Exp1}

The study area $A$ was a square starting at (0,0) and 10 units on each side. Data were simulated using rejection sampling.  Consider the intensity surface $\lambda(x,y) = x/20 + y/20$, which increases linearly from 0 at (0,0) to 1 at (10,10) within $A$.  A location was simulated with $x^*$, $y^*$, and $z^*$, where each was drawn from $\Unif(0,1)$. If $z^* < \lambda(x^*,y^*)$, then the simulated location was retained.  The set $(x^*,y^*,z^*)$ was drawn 2000 times, so the expected number of locations retained was 1000 per simulated data set, but note that the actual number varied randomly among simulations.  For each simulated data set, sample units as square plots that measured 0.3 on a side were systematically placed in a $16 \times 16$ grid as shown in Figure~\ref{fig:SimulationPlots}.  The 256 plots covered 23.04\% of the study area $A$.

The results of experiment 1 are given in Table~\ref{Tab:Sim1}. Note that for SRS and various knot proportions, there was little bias, which is $<$ 1\% of the average total.  All methods had very similar RMSPE as well, and 90\% confidence intervals generally had the appropriate coverage, no matter which overdispersion method was used.  Of course, the data were not simulated with overdispersion.  The only exception occurred when using many knots for the trimmed overdispersion estimators CI90$_{TL}$ and CI90$_{TG}$, where variance was overestimated.


\subsection{Experiment 2: Inhomogeneous spatial point process with irregular sampling} \label{sec:Exp2}

For simulation experiment 2, locations were simulated exactly as they were in Experiment 1 (Section~\ref{sec:Exp1}).  However, we created unbalanced spatial sampling by removing one column and two rows of sample units, as shown in Figure~\ref{fig:SimulationPlots}.  In this case, 210 plots covered 18.9\% of the study area $A$.

The results of experiment 2 are given in Table~\ref{Tab:Sim2}, which should not be surprising for SRS.  Indeed, the goal of this research was to find good estimators when high (or low) abundance areas were oversampled, and SRS makes no weighting adjustments for this.  Consequently, it had a large bias, whereas $\widehat{T}(A)$ in (\ref{eq:ThatA}) remained relatively unbiased with RMSPE only slightly larger than experiment 1 (note, too, that sample sizes were smaller here).  The same basic patterns appeared for CI90, with generally valid confidence intervals (except for SRS), except when many knots are used for the trimmed overdispersion estimators CI90$_{TL}$ and CI90$_{TG}$.


\subsection{Experiment 3: Double spatial cluster process on inner rectangle} \label{sec:Exp3}

One difficult situation for estimating abundance from counts occurs when there is a large number of zeros and there is fine scale clustering, so we tested the abundance estimator under both of those conditions. For this experiment, seed points were simulated within a rectangle within the study area. Again, the study area $A$ was a square starting at (0,0) and 10 units on each side.  The inner rectangle itself had random boundaries, where the lower x-axis and y-axis boundaries were each randomly drawn from $\Unif(3.5,4.5)$, and the upper x-axis and y-axis boundaries were each drawn from $\Unif(7.5,8.5$). Next 100 parent seed points were uniformly simulated over the inner rectangle (such a rectangle is shown with solid lines in Figure~\ref{fig:SimulationExample}), and then each parent had a random number, $\Poi(15)$, of children that were uniformly distributed on a square with sides of length 2 centered on each parent.  A second finer scale cluster process was added by creating 25 more parent points uniformly distributed over the inner rectangle, where each parent had a random number, $\Poi(9)$, of children that were uniformly distributed on a square with sides 0.4 centered on each parent.  After simulating all points, they were thinned using the same function as experiments 1 and 2, by simulating $z^*_i \sim \Unif(0,1)$ at each simulated location with coordinates $x_i$ and $y_i$, and keeping that location if $z^*_i < x_i/20 + y_i/20$. From 1000 simulated experiments, this yielded an average of 1034 points per simulation.  An example of one simulation is given in Figure~\ref{fig:SimulationPlots}.  The sample units were placed in the same positions as for Experiment 2 (Section~\ref{sec:Exp2}).

The results of experiment 3 are given in Table~\ref{Tab:Sim3}. Once again, because of the unbalanced sampling, SRS had a large RMSPE and was highly biased, whereas $\widehat{T}(A)$ in (\ref{eq:ThatA}) remained unbiased at generally less than 0.5\% of the total. RMSPE was larger than experiments 1 and 2, but this is not surprising given the smaller area with positive count values.  This experiment showed some poorer performance for CI90, especially for  CI90$_{OD}$, which underestimated variance with coverage nearer 80\% rather than 90\%. Both CI90$_{WR}$ and $CI90_{TG}$ performed more poorly with increasing numbers of knots. CI90$_{TL}$ coverage is about 3\% low for the fewest number of knots, but generally improves with the number of knots. Notice also that there is almost a 2\% chance that the parameter estimation algorithm will fail when there are many knots.


\subsection{Experiment 4: Double spatial cluster process on double inner rectangles} \label{sec:Exp4}

\begin{figure}[H]
	\begin{center}
	\includegraphics[width=450pt]{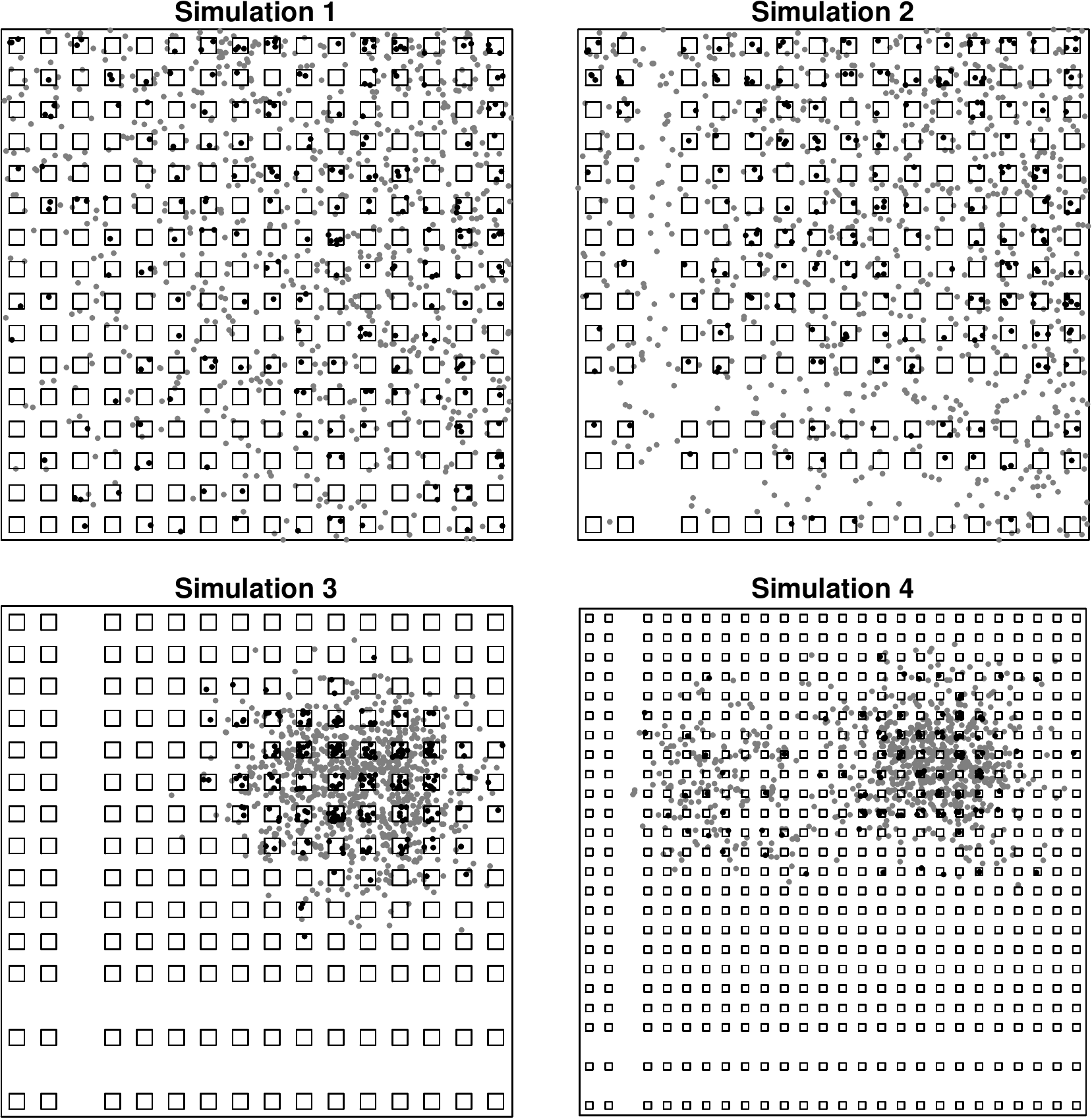}
	\end{center}
	\caption{Examples of simulated data used to test methods.  All simulated points are shown as grey and black dots.  Sample units are shown as squares, and black dots were in sample units while the grey dots were out. The four types of simulations are described in the text.\label{fig:SimulationPlots}}
	\end{figure}

For this experiment, the seed points were simulated in a manner similar to experiment 3.  However, the sample units were made smaller, with length 0.14 on a side, on a 26 $\times$ 26 grid in a study area, $A$, which was again a square starting at (0,0) and 10 units on each side.  This time there were two inner rectangles. One inner rectangle had random boundaries where the lower x-axis and y-axis boundaries were each randomly drawn from $\Unif(5.8,6.2)$, and the upper x-axis and y-axis boundaries were each drawn from $\Unif(7.8,8.2$).  Next 75 parent seeds were uniformly simulated over this rectangle, and each parent seed had a random number, $\Poi(14)$, of children uniformly distributed in a box with sides of length 2 centered on each parent. A finer scale cluster process was added by creating 25 more parent points uniformly distributed over this inner rectangle, where each parent had a random number, $\Poi(8)$, children that were uniformly distributed on a square with sides 0.4 centered on each parent. A second inner rectangle had random boundaries where the lower x-axis boundary was randomly drawn from $\Unif(0.8,1.2)$ and the upper x-axis boundary was drawn from $\Unif(3.8,4.2)$, while the lower y-axis boundary was randomly drawn from $\Unif(4.8,5.2)$, and the upper y-axis boundary was drawn from $\Unif(7.8,8.2$).  Here, 25 parent seeds were uniformly simulated over this rectangle, and each parent seed had a random number, $\Poi(14)$, of children uniformly distributed in a box with sides of length 1 centered on each parent. A finer scale cluster process was achieved by creating 10 more parent points uniformly distributed over this inner rectangle, where each parent had a random number, $\Poi(8)$, of children that were uniformly distributed on a square with sides 0.4 centered on each parent. After simulating all points, they were thinned using the same function as experiments 1-3, by simulating $z^*_i \sim \Unif(0,1)$ at each simulated location with coordinates $x_i$ and $y_i$, and keeping that location if $z^*_i < x_i/20 + y_i/20$. From 1000 simulated experiments, this yielded an average of 1012 points per simulation. An example of one simulation is given in Figure~\ref{fig:SimulationPlots}. We created unbalanced spatial sampling by removing one column and two rows of sample units, as shown in Figure~\ref{fig:SimulationPlots}. In this case, 600 plots covered 11.8\% of the study area $A$.

The results of experiment 4 are given in Table~\ref{Tab:Sim4}. Once again, because of the unbalanced sampling, SRS had a large RMSPE and was highly biased, whereas $\widehat{T}(A)$ in (\ref{eq:ThatA}) remained unbiased for smaller number of knots, but was $>$ 1\% of the total for the largest number of knots. RMSPE was less than experiments 3, likely due to a larger area with nonzero counts and smaller plots, leading to a larger sample size.  Once again, CI90$_{OD}$ underestimated variance with coverage nearer 83\% rather than 90\%. CI90$_{WR}$ was about 4\% high for small number of knots, but improved with numbers of knots. CI90$_{TG}$ remains about 2\% high for all combinations of knot numbers, while CI90$_{TL}$ is 1\% to 2\% low for all combinations of knot numbers. Here, notice also that there is over 24\% chance that the parameter estimation algorithm will fail when there are many knots.

\begin{figure}[H]
	\begin{center}
	\includegraphics[width = .8\maxwidth]{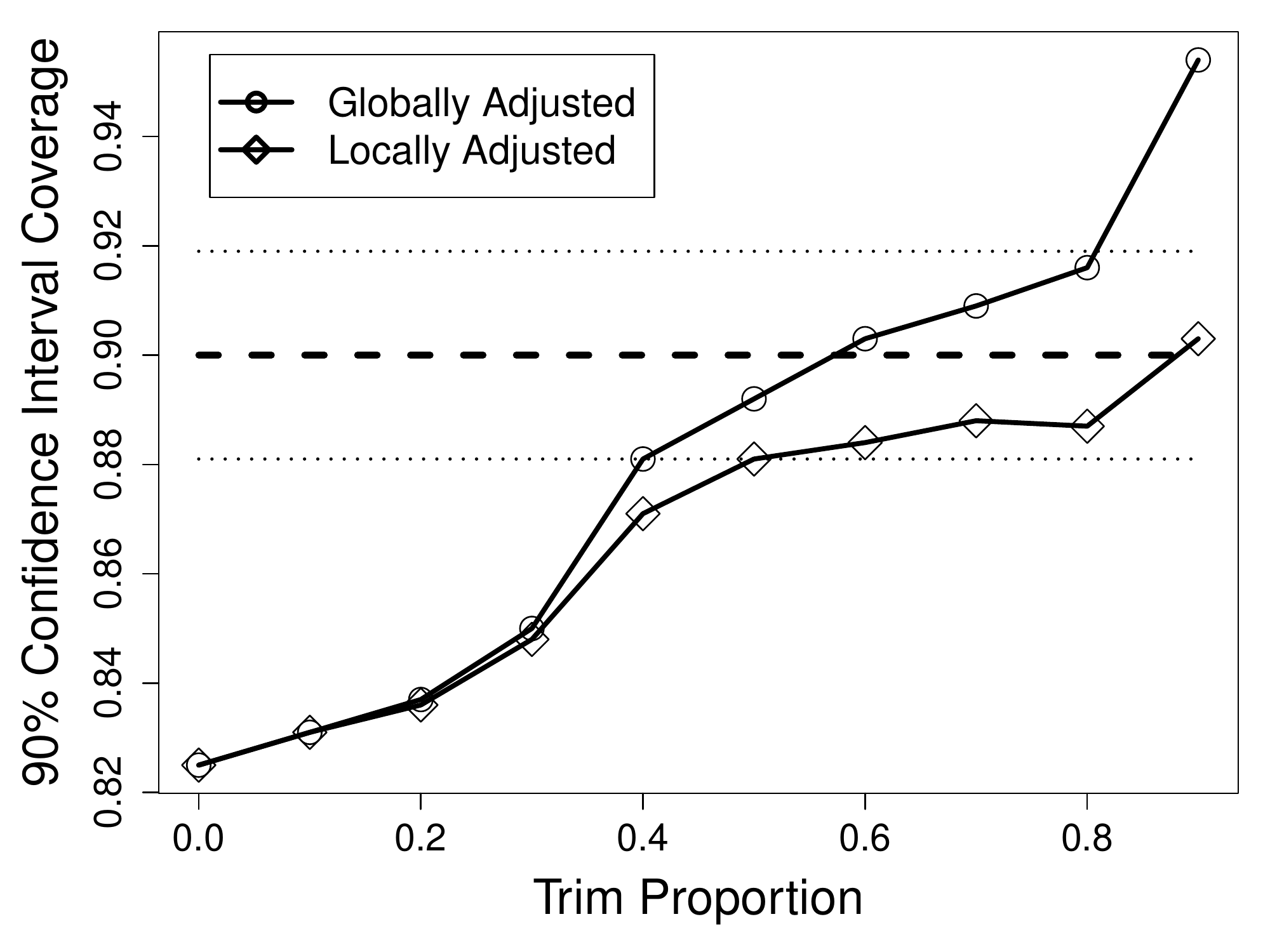}
	\end{center}
	\caption{The 90\% confidence interval coverage for various overdispersion trim proportions, using $K_C=5$ and $K_F=16$ for 1000 simulations from simulation experiment 4.  The 90 \% line is shown as a dashed horizontal line, and the dotted horizontal lines show the 95\% bounds for an estimator that had a true coverage of 0.90.  \label{EffectTrimProp}}
	\end{figure}

The effects of varying $p$ in CI90$_{TG}$ and CI90$_{TL}$ are shown in Figure~\ref{EffectTrimProp}, using 1000 simulated data sets as described for experiment 4. Notice that confidence coverage for CI90$_{TG}$ was in the ``valid'' zone  between $p = 0.4$ and $p = 0.8$, and CI90$_{TL}$ was in the ``valid'' zone  when $p \ge 0.5$. In a real setting, such a $p$ value must be chosen by data examination (at least without further research).  Looking at the simulated data in Figure~\ref{fig:SimulationPlots}, one would try to estimate the area dominated by zero counts, and then trim them.  A $p$ value anywhere from 0.6 to 0.8 seems reasonable, and would lead to nearly correct confidence intervals using either CI90$_{TG}$ or CI90$_{TL}$.  Clearly, CI90$_{TG}$ is a more conservative strategy, but trimming aggressively with CI90$_{TL}$ appears viable. Another consideration is that we expect that smaller $p$ values would be more efficient by using more data.


\section{Example: aerial surveys of harbor seals} \label{sec:Example}

\begin{figure}[H]
	\begin{center}
	\includegraphics[height=350pt]{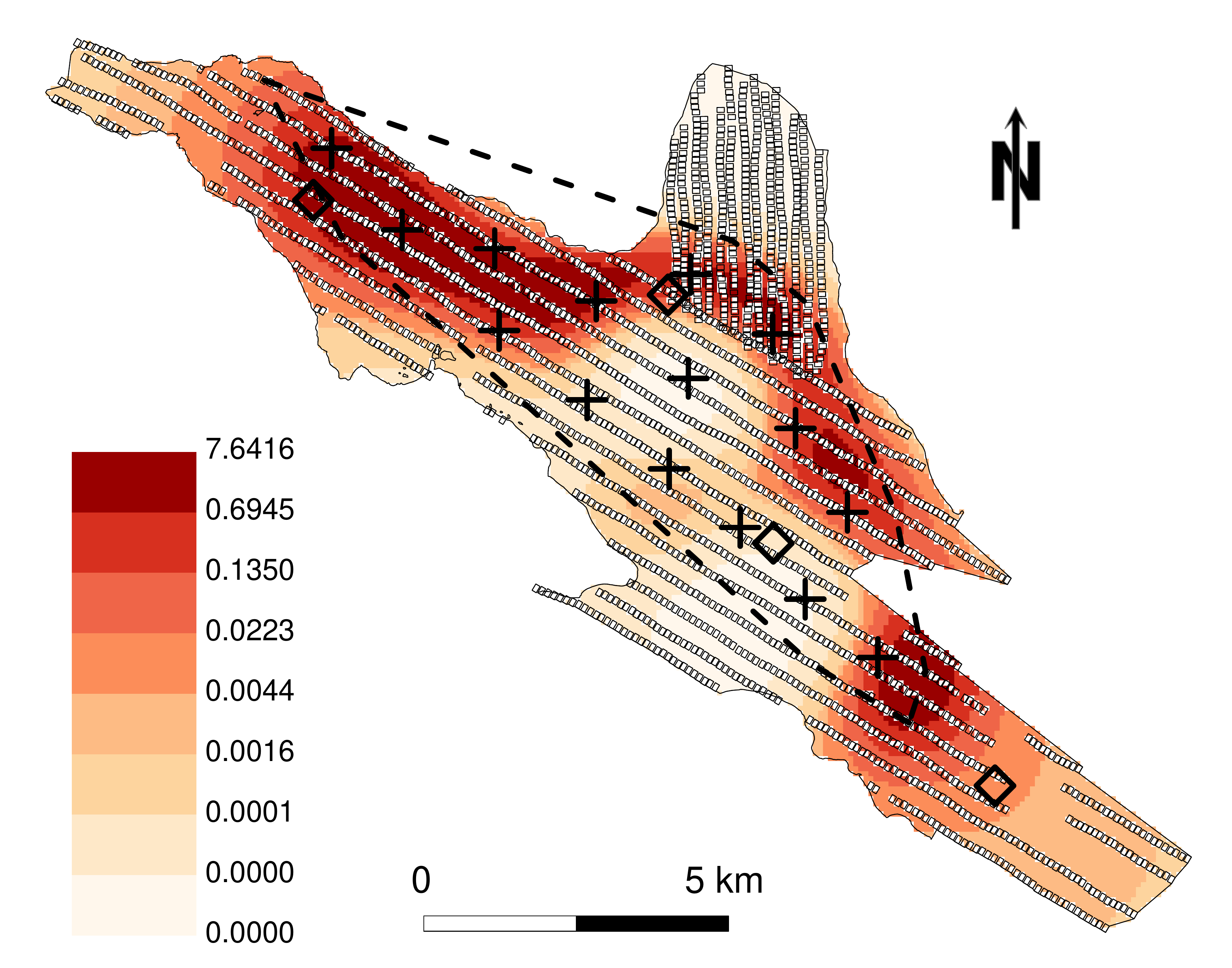}
	\end{center}
	\caption{The study area for the harbor seal data with the fitted intensity surface throughout $\cU$, scaled to yield the expected count per prediction grid block. The coarse scale knots are shown as open diamonds while the fine scale knots are shown as crosses. The minimum convex polygon enclosing all plots with nonzero counts is given by the dashed line. \label{RealDataFittedSurface}}
	\end{figure}

The study area boundary, the locations of all 2080 plots, and the observed counts of seals within plots are shown in Figure~\ref{fig:studyArea}, and the data were summarized in Section~\ref{sec:motivate}.  We used $K_C$ = 4 knots in the coarse grid and we used $K_F$ = 15 knots in the fine grid shown in Figure~\ref{RealDataFittedSurface}. Notice that the fine grid of knots is contained in a bounding rectangle around only those plots with non-zero counts; with the reason explained in Section~\ref{sec:EvalExp}. The model fit took 17.56 seconds on a Intel Xeon 2.66GHz processor running under the linux operating system. The estimate range parameter $\rho_F$ was 1.81 km, and $\rho_C$ was 4.04 km. The fitted intensity surface is shown in Figure~\ref{RealDataFittedSurface}. The estimate obtained from integrating this surface, along with the observed count, using $\widehat{T}(A)$ in (\ref{eq:ThatA}) was 4012.  The standard error $\sqrt{\hat{\cM}(\hat{T}(A))}$ in (\ref{eq:nu1}), without any corrections for overdispersion, was 111.86.  Interestingly, the $\hat{\mu}(\cU)$ component of (\ref{eq:nu1}) was 3010, and the $\bc^\prime\hat{\bSigma}\bc$ component of (\ref{eq:nu1}) was 9504.  The estimates of overdispersion given in Section~\ref{sec:overdispersion} were $\omega_{OD}$ = 17.26, $\omega_{WR}$ = 3.77, $\omega_{TG}$ = 3.45, and $\omega_{TL}$ = 2.79. If we use $\omega_{TG}$ for overdispersion, that yields a standard error of 386.  Then the 95\% confidence interval, from (\ref{eq:CI}), for the estimate of total abundance is (3322, 4845).

We tried different combinations of knots and $p$ in $\omega_{TG}$ and $\omega_{TL}$.  No systematic attempt is made to present those results, and knot selection and overdispersion is a topic for future research.  However, we did note that the abundance estimate was little changed for knots up to $K_C = 8$ and $K_F = 32$.  However, there were changes in overdispersion estimates. Also, we noticed that for the knots that we selected, increasing $p$ in $\omega_{TG}$ and $\omega_{TL}$ \emph{decreased} overdispersion, rather than the increasing values seen in Figure~\ref{EffectTrimProp}. Conceptually, it is possible that an area with high counts could have less variability than a larger area that includes both high counts and low counts.  Also, note that $\omega_{OD}$ = 17.26, which is much larger than one, and larger than all other overdispersion estimators, in contrast to the example provided by Figure~\ref{fig:residDisp}. However, the explanation is also provided by Figure~\ref{fig:residDisp} because for this data set there were a few non-zero counts with very low expected values that dominated $\omega_{OD}$.  We chose this example in part because it showed some exceptions for the overdispersion parameters.  Based on dozens of similar real examples, most of them have $\omega_{OD} = 1$, and this appears to be an unstable estimator for our purposes.

\section{Discussion and conclusions} \label{sec:DiscConc}

Our objectives were to develop a model based estimator of a total by using counts from irregularly spaced plots.  We wanted this estimator to have goals, properties, and performance similar to classical sampling: to estimate the realized total, not the mean of an assumed process, to have a finite population correction factor, to be unbiased with valid confidence intervals that must be robust to nonstationary mean and variance, and to be fast to compute. The problem was made more difficult by features of the data that were seen in the real example.  First, the nonzero counts were highly clustered in space, and there were large areas of zeros.   Secondly, the counts, where they occurred, showed overdispersion.  Finally, sample sizes were quite large, so we needed to use computationally efficient methods.  The general approach that we took was to assume that the data came from an inhomogeneous point process with overdispersion. We modeled the intensity surface of the inhomogeneous point process using spatial basis functions in a generalized linear mixed model framework.  

To test the method, we started with fairly benign conditions in experiment 1.  We then added complexity to the simulations that matched the complexity seen in the real data, by simulating spatially unbalanced sampling, data with overall trend in point density, several areas of clustered points, overdispersion within the cluster areas, and yet large areas with zeros, culminating in experiment 4.  Overall, our method worked well, especially when using some of the newly introduced overdisperion estimators. One of the main contributions of this manuscript was the introduction of overdispersion estimators when the overdispersion appears to be varying spatially (a nonstationary overdispersion).  

One of the interesting findings from our research was the effect of knot placement and proportion (Tables~\ref{Tab:Sim1} - \ref{Tab:Sim4}). Initially, we found a lack of convergence when fitting these models if there were too many knots, with short ranges, over areas containing all zeros. In retrospect, that may not seem surprising, but we have not found it reported in the literature.  For that reason, we created spatially restricted knots over areas with non-zero values (dashed line in Figure~\ref{fig:SimulationExample}; also see Figure~\ref{RealDataFittedSurface}). For placing knots, we used a K-means clustering algorithm on the spatial coordinates to create a space-filling design. Other approaches that could be used were mentioned in Section~\ref{sec:Knots}. Our results show that the bias does not depend very much on the number of knots, but the standard errors are quite sensitive.  An encouraging result is that standard errors yield better confidence intervals when the number of knots is quite small, making the algorithm very fast. The whole issue of knot selection needs further research.

Besides more research on overdispersion estimators, and knot placement and number, there are numerous modifications that could be applied to our basic approach. We chose spatial basis functions that were Gaussian kernels at two scales, and there are many obvious modifications.  Because we annually analyze dozens of data sets like our example for harbor seals, we took a maximum likelihood approach to estimating parameters and total abundance; however, Bayesian approaches could also be used \citep{Wikl:spat:2002, Chri:Waag:baye:2002}.  The ability to estimate the intensity function, essentially point-level information, from data at an aggregated level, i.e., counts from plots, depends on the plots being small in relation to changes in the intensity function. If plots are very large, then methods in this paper could still be used, but spatial points would need to be mapped within plots, and more traditional methods from the spatial point process literature could be used for estimating the intensity surface. For example, the R \citep{R:Deve:Core:ALan:2014} package spatstat \citep{Badd:Turn:spat:2005} can be used when the realized point patterns have a complicated ``mask'' comprised of many disjoint plots \citep{Badd:Turn:mode:2006}. Alternatively, area-to-point geostatistical methods \citep{Kyri:geos:2004} could be used. The main point is that once the intensity surface is estimated, (\ref{eq:ThatA}) can be used to estimate total abundance.  The variance of the total abundance can be estimated with (\ref{eq:nu1}) if maximum likelihood methods are used, along with one of the overdispersion factors (Section~\ref{sec:overdispersion}) that make sense for the problem under consideration.

 
\section*{ACKNOWLEDGMENTS} 

The project received financial support from NOAA’s National Marine Fisheries Service, Alaska Fisheries Science Center. In kind support was provided by the Yakutat office of the U.S. National Weather Service and the Yakutat Tlingit Tribe whose concerns about the seal population provided the impetus for this study.  Planes were provided by NOAA's Office of Marine and Aviation Operations.  We appreciate the safety awareness and skill exhibited by the NOAA Corp pilots in planning for and completing surveys in very challenging conditions.  The findings and conclusions in this paper are those of the authors and do not necessarily represent the views of the National Marine Fisheries Service.



\clearpage

\section*{TABLES}


\footnotesize
\begin{table}[ht]
\caption{Results for bias, RMSPE, confidence interval coverage, and failure rate for simulation experiment 1. The number of coarse-scale knots used is given by $K_C$, and $K_F$ is the number of fine-scale knots. An example of a single simulated data set is given in Figure~\ref{fig:SimulationPlots}. \label{Tab:Sim1}}
\begin{center}
\begin{tabular}{rrrrrr}
  \hline
  \hline
	&  & \multicolumn{4}{c}{Knots} \\ 
 & SRS & $\begin{array}{c}K_C=3  \\ K_F=8 \end{array}$ & $\begin{array}{c}K_C=5  \\ K_F=16 \end{array}$ & $\begin{array}{c}K_C=7  \\ K_F=24 \end{array}$ & $\begin{array}{c}K_C=9  \\ K_F=32 \end{array}$ \\
  \hline
 Bias & 6.425 & -1.277 & -9.735 & 7.048 & 5.941 \\ 
  RMSPE & 58.060 & 57.493 & 59.036 & 58.243 & 58.038 \\ 
  CI90$^a$ & 0.914 & 0.892 & 0.886 & 0.886 & 0.893 \\ 
  CI90$_{OD}^b$ &  & 0.917 & 0.921 & 0.890 & 0.900 \\ 
  CI90$_{WR}^c$ &  & 0.895 & 0.890 & 0.886 & 0.894 \\ 
  CI90$_{TG}^d$ &  & 0.909 & 0.922 & 0.936 & 0.958 \\ 
  CI90$_{TL}^e$ &  & 0.901 & 0.898 & 0.911 & 0.928 \\ 
  Fail Rate$^f$ & 0.000 & 0.000 & 0.000 & 0.000 & 0.000 \\

   \hline
\end{tabular}
\end{center}
\textrm{$^a$ 90 \% confidence interval coverage based on standard errors without overdispersion.} \\ 
\textrm{$^b$ 90 \% confidence interval coverage using classical overdispersion} \\
\textrm{$^c$ 90 \% confidence interval coverage using a weighted regression overdisperion estimator} \\
\textrm{$^d$ 90 \% confidence interval coverage using a global trimmed mean overdisperion estimator} \\
\textrm{$^e$ 90 \% confidence interval coverage using a local trimmed mean overdisperion estimator} \\
\textrm{$^f$ failure of the estimator due to lack of convergence or excessively large estimates or standard errors} \\
\end{table}


\footnotesize
\begin{table}[ht]
  \caption{Results for bias, RMSPE, confidence interval coverage, and failure rate for simulation experiment 2. Details on column and row labels are given in Table~\ref{Tab:Sim1} and the text. An example of a single simulated data set is given in Figure~\ref{fig:SimulationPlots}. Row names are described in Table~\ref{Tab:Sim1}. \label{Tab:Sim2}}
\begin{center}
\begin{tabular}{rrrrrr}
  \hline
  \hline
	&  & \multicolumn{4}{c}{Knots} \\ 
 & SRS & $\begin{array}{c}K_C=3  \\ K_F=8 \end{array}$ & $\begin{array}{c}K_C=5  \\ K_F=16 \end{array}$ & $\begin{array}{c}K_C=7  \\ K_F=24 \end{array}$ & $\begin{array}{c}K_C=9  \\ K_F=32 \end{array}$ \\
  \hline
 Bias & 79.234 & -1.333 & -7.632 & 14.511 & 13.856 \\ 
  RMSPE & 104.979 & 66.347 & 68.846 & 68.311 & 68.527 \\ 
  CI90$^a$ & 0.726 & 0.883 & 0.864 & 0.883 & 0.883 \\ 
  CI90$_{OD}^b$ &  & 0.913 & 0.927 & 0.894 & 0.892 \\ 
  CI90$_{WR}^c$ &  & 0.889 & 0.874 & 0.888 & 0.886 \\ 
  CI90$_{TG}^d$ &  & 0.901 & 0.913 & 0.952 & 0.973 \\ 
  CI90$_{TL}^e$ &  & 0.893 & 0.887 & 0.919 & 0.939 \\ 
  Fail Rate$^f$ & 0.000 & 0.000 & 0.000 & 0.000 & 0.001 \\

   \hline
\end{tabular}
\end{center}
\end{table}


\footnotesize
\begin{table}[ht]
\caption{Results for bias, RMSPE, confidence interval coverage, and failure rate for simulation experiment 3. Details on column and row labels are given in Table~\ref{Tab:Sim1} and the text. An example of a single simulated data set is given in Figure~\ref{fig:SimulationPlots}. Row names are described in Table~\ref{Tab:Sim1}. \label{Tab:Sim3}}
\begin{center}
\begin{tabular}{rrrrrr}
  \hline
  \hline
	&  & \multicolumn{4}{c}{Knots} \\ 
 & SRS & $\begin{array}{c}K_C=3  \\ K_F=8 \end{array}$ & $\begin{array}{c}K_C=5  \\ K_F=16 \end{array}$ & $\begin{array}{c}K_C=7  \\ K_F=24 \end{array}$ & $\begin{array}{c}K_C=9  \\ K_F=32 \end{array}$ \\
  \hline
 Bias & 214.816 & -2.389 & -4.365 & -2.919 & -1.637 \\ 
  RMSPE & 235.713 & 79.207 & 79.250 & 79.285 & 80.175 \\ 
  CI90$^a$ & 0.774 & 0.780 & 0.777 & 0.780 & 0.783 \\ 
  CI90$_{OD}^b$ &  & 0.807 & 0.790 & 0.788 & 0.787 \\ 
  CI90$_{WR}^c$ &  & 0.913 & 0.903 & 0.864 & 0.848 \\ 
  CI90$_{TG}^d$ &  & 0.930 & 0.935 & 0.938 & 0.947 \\ 
  CI90$_{TL}^e$ &  & 0.877 & 0.877 & 0.872 & 0.901 \\ 
  Fail Rate$^f$ & 0.000 & 0.000 & 0.000 & 0.000 & 0.018 \\

   \hline
\end{tabular}
\end{center}
\end{table}


\footnotesize
\begin{table}[ht]
\caption{Results for bias, RMSPE, confidence interval coverage, and failure rate for simulation experiment 4. Details on column and row labels are given in Table~\ref{Tab:Sim1} and the text. An example of a single simulated data set is given in Figure~\ref{fig:SimulationPlots}.  Row names are described in Table~\ref{Tab:Sim1}. \label{Tab:Sim4}}
\begin{center}
\begin{tabular}{rrrrrr}
  \hline
  \hline
	&  & \multicolumn{4}{c}{Knots} \\ 
 & SRS & $\begin{array}{c}K_C=3  \\ K_F=8 \end{array}$ & $\begin{array}{c}K_C=5  \\ K_F=16 \end{array}$ & $\begin{array}{c}K_C=7  \\ K_F=24 \end{array}$ & $\begin{array}{c}K_C=9  \\ K_F=32 \end{array}$ \\
  \hline
 Bias & 148.523 & 5.179 & 3.440 & 7.287 & 14.629 \\ 
  RMSPE & 163.516 & 60.403 & 61.021 & 62.102 & 64.136 \\ 
  CI90$^a$ & 0.834 & 0.837 & 0.828 & 0.831 & 0.827 \\ 
  CI90$_{OD}^b$ &  & 0.847 & 0.831 & 0.833 & 0.828 \\ 
  CI90$_{WR}^c$ &  & 0.937 & 0.926 & 0.917 & 0.910 \\ 
  CI90$_{TG}^d$ &  & 0.920 & 0.916 & 0.905 & 0.906 \\ 
  CI90$_{TL}^e$ &  & 0.892 & 0.892 & 0.878 & 0.867 \\ 
  Fail Rate$^f$ & 0.000 & 0.000 & 0.000 & 0.000 & 0.242 \\

   \hline
\end{tabular}
\end{center}
\end{table}


\clearpage

\section*{FIGURES}


	%
	%


	%
	%


	%
	%


		%
		%

	%
	%
	%


	%
	%
	%


	%
	%
	%


\end{document}